\documentclass[structabstract]{aa}  
\usepackage{graphicx}
\usepackage{natbib}
\usepackage{txfonts}
\begin{document}
\def\frac{$''$\hspace*{-.1cm}}
\def\deg{$^{\circ}$}
\def\min{$'$}
\def\deg{$^{\circ}$\hspace*{-.1cm}}
\def\hii{H\,{\sc ii}}
\def\hi{H\,{\sc i}}
\def\hg{H$\gamma$}
\def\hd{H$\delta$}
\def\he{H$\epsilon$}
\def\hb{H$\beta$}
\def\ha{H$\alpha$}

\def\lam{$\lambda$}

\def\aiii{Al~{\sc{iii}}\ }
\def\ariii{[Ar\,{\sc{iii}}]}
\def\ariv{[Ar\,{\sc{iv}}]}
\def\cii{[C\,{\sc{ii}}]}
\def\ciii{C\,{\sc{iii}}}
\def\civ{C\,{\sc{iv}}}
\def\caii{Ca\,{\sc{ii}}}
\def\cliii{[Cl\,{\sc{iii}}]}
\def\crii{Cr\,{\sc{ii}}}
\def\fei{Fe\,{\sc{i}}}
\def\feii{Fe\,{\sc{ii}}}
\def\feiii{Fe\,{\sc{iii}}}
\def\hei{He\,{\sc{i}}}
\def\heii{He\,{{\sc ii}}}
\def\nii{[N\,{\sc{ii}}]}
\def\niii{N\,{\sc{iii}}}
\def\niv{N\,{\sc{iv}}}
\def\nv{N\,{\sc{v}}}
\def\ni{Na~{\sc{i}}\ }
\def\nei{Ne~{\sc{i}}\ }
\def\neiii{[Ne~{\sc{iii}}]}
\def\oi{[O\,{\sc i}]}
\def\oii{[O\,{\sc ii}]}
\def\oiii{[O\,{\sc iii}]}
\def\sii{[S\,{\sc ii}]}
\def\siii{[S\,{\sc iii}]}
\def\siv{S\,{\sc{iv}}}
\def\si_ii{Si\,{\sc{ii}}}
\def\si_iii{Si\,{\sc{iii}}}
\def\siiv{Si\,{\sc{iv}}}
\def\srii{Sr\,{\sc ii}}
\def\tiii{Ti\,{\sc{ii}}}
\def\yii{Y\,{\sc ii}}

\def\x{$\times$}
\def\av{$A_{V}$}
\def\mcube{$^{-3}$}
\def\cm2{cm$^{-2}$}
\def\sec{s$^{-1}$}

\def\sm{$M_{\odot}$}
\def\slum{$L_{\odot}$}
\def\ab{$\sim$}
\def\sec{s$^{-1}$}

\title{An interesting candidate for isolated massive-star formation in the Small Magellanic Cloud
\thanks{Based on observations obtained at the European Southern 
   Observatory, La Silla, Chile, Program 69.C-0286(A) and 69.C-0286(B).}}

\author{R. Selier\inst{1} \and M. Heydari-Malayeri\inst{1} \and
           D. A. Gouliermis\inst{2}
}

\institute{Laboratoire d'Etudes du Rayonnement et de la Mati\`ere en Astrophysique (LERMA), 
Observatoire de Paris, CNRS, \\ 
61 Avenue de l'Observatoire, 75014 Paris, France,  Romain.Selier@obspm.fr
\and
Max-Planck-Institut f\"{u}r Astronomie, K\"{o}nigstuhl 17, 69117 Heidelberg, Germany
}

\authorrunning{Selier et al.}
\titlerunning{SMC compact \hii\ region N33}

\date{Received 8 November 2010 / accepted 3 February 2011}

 
   \abstract
   {The region of the Small Magellanic Cloud (SMC) with which this paper is concerned 
contains the highest concentration of IRAS/Spitzer sources, \hi\ emission, and 
molecular clouds in this neighboring galaxy. However, it has been the target of very 
few studies, despite this evidence of star formation. }
  {We present the first detailed study of the compact \hii\ region N33 in the SMC by placing it in 
a wider context of massive star formation. Moreover, we show that N33 is a particularly 
interesting candidate for isolated massive star formation.
}
   {This analysis is based mainly on optical ESO NTT observations,  
both imaging and spectroscopy, coupled with other archive data, notably Spitzer images 
(IRAC 3.6, 4.5, 5.8, and 8.0\,$\mu$m) and  2MASS observations.
} 
   {We derive a number of physical characteristics of the compact \hii\ region N33 
for the first time. This gas and dust formation of 7\frac.4 (2.2 pc) in diameter 
is powered by a massive star of spectral type 
O6.5-O7\,V.  The compact \hii\ region belongs to a rare class of \hii\ regions in the 
Magellanic Clouds, called high-excitation blobs (HEBs). 
We show that this \hii\ region is not related to any star  
cluster. Specifically, we do not find any traces of clustering around N33 on 
scales larger than 10\frac\, (\ab\,3 pc).  On smaller scales, there is a 
marginal stellar concentration, the low density of which, below the 3$\sigma$ level, 
does not classify it as a real cluster. We also verify 
that N33 is not a member of any large stellar association. Under these circumstances, 
N33 is also therefore attractive because 
it represents a remarkable case of isolated massive-star formation in the SMC. Various 
aspects of the relevance of N33 to the topic of massive-star formation in isolation are discussed.

}
   {}


   \maketitle
%

\section{Introduction}

The Small Magellanic Cloud (SMC) provides a unique opportunity for studying 
star formation in low-metallicity environments. Although a neighboring galaxy, 
the SMC is considered to be reminiscent of high-redshift galaxies. It has therefore been the 
subject of numerous surveys and studies at various wavelengths. However, as far as massive 
star formation is concerned, detailed study of its individual star-forming regions  
has focused on a relatively small number of cases. Apart from the pre-eminent region 
N66, to which a considerable amount of research work has been devoted 
\citep[][and references therein]{MHM10a}, only several star-forming regions 
associated with the Shapley's wing \citep[][]{Shapley40}, such as 
N81 \citep[][]{MHM99a}, 
N88A \citep[][]{MHM99b}, 
N83/N84 \citep[][]{Lee09}, and  
N90 \citep[][]{Cignoni09} 
have been studied in more detail.  \\

In particular, the star-forming regions populating the lower portion of the SMC bar, 
situated south-west of N66, have so far received little attention. 
The southern end of the SMC has the particularity of containing the highest concentration 
of molecular clouds, IRAS/Spitzer sources, and \hi\ emission in this neighboring galaxy  
\citep[][]{Fukui10,Mizuno01,Stanimirovic00,Stanimirovic99,Blitz07,Bolatto07,Leroy07,Bot10}. 
However, higher spatial resolutions and more extended mappings, 
especially in CO observations,  
are required to precisely associate the molecular clouds of that part with 
individual \hii\ regions and study the interaction with massive stars.  
The \hii\ region  LHA\,115-N33, or in short N33 \citep[][]{Henize56}, 
with J2000.0 equatorial coordinates $\alpha$\,=\,00h\,49m\,29s and 
$\delta$\,=\,-73\deg\,\,26\min\,\,34\frac , belongs to this part of the SMC. 
Very few studies have been devoted to it, despite its several interesting 
characteristics, which the present paper aims to highlight. In particular, 
as the present paper will show,
this object belongs to a small class of Magellanic compact \hii\ regions, called 
high-excitation blobs \citep[HEBs; for a review see][]{MHM10b}.
N33 appears also under number 138 in the catalog of emission-line stars and 
planetary nebulae compiled by \citet[][]{Lindsay61}. In contrast, it  
is not identified in the optical survey of \citet[][]{DEM76}.
Likewise, N33 bears number 297 in the catalog of \citet[][]{Azzo93} among the list 
of proven or probable SMC compact or small \hii\ regions. 
N33 was detected as the IRAS source 00477-7343 
\citep[][]{Helou88}, and in far-IR as the source \#28 in the ISO 170 $\mu$m catalog 
\citep[][]{Wilke03}.  
N33 was also included in several radio continuum surveys, 
including the Australia Telescope Compact Array (ATCA) observations at 
3 and 6 cm \citep[][]{Filipovic98,Indebetouw04}. \\

This region of the SMC also hosts a planetary nebula. This was first suspected by  
\citet[][]{Henize63}, who nevertheless could not confirm its presence. Subsequently, 
\citet[]{Jacoby02} detected a planetary nebula candidate at the position (J2000.0) 
$\alpha$\,=\,00h\,49m\,35.0s and $\delta$\,=\,-73\deg\,\,26\min\,\,36\frac.9 
(their object number 7). This means that the candidate planetary nebula should lie  
27\frac\, east of the \hii\, region. The reported diameter of the planetary nebula 
is  7\frac\, \citep[]{Jacoby02}. Although this planetary nebula is probably 
unrelated to the \hii\, region N33, it has created some confusion. In the 
CDS Simbad database, N33 appears under the heading of planetary nebula. 
It is also notable that among the 11 references given (covering the 1956-2008 period),  
only two are directly concerned with the planetary nebula candidate.   \\ 

As a notable point, this object does not belong to any known stellar association 
of the SMC \citep[][]{Hodge85}. In the context of isolated massive star formation 
studies, this makes N33 particularly interesting. We attempt to show that N33 
is  an isolated \hii\ region powered by 
a massive star, which has probably formed in isolation. By isolation, we mean not 
being traceable to an origin in any OB association. 
Observational findings suggest that massive stars generally form in groups 
\citep[e.g.,][]{Zinnecker07,Schilbach08,Gvaramadze08}.  
There is, however, a statistically small percentage of massive stars 
(\ab\,5\%) that form in isolation \citep[][]{deWit05,Parker07}. 
It is not yet understood which physical conditions favor the formation 
of isolated massive stars and how they form. It is 
pertinent to know whether the same continuous process governs the formation of 
clusters of massive stars in large molecular clouds and single stars in 
smaller molecular clouds \citep[see, e.g.,][and below Sect. 4]{Oey04}. 
Should the answer be negative, it would be necessary to know  
the mechanisms responsible for the formation of 
massive stars in isolation. We discuss several aspects 
of this topic.  The study of this family 
of massive stars would provide new insights into a clearer 
understanding of massive 
stars in general, which is not entirely clear. Apart from studying global 
aspects of isolated massive star formation, it is necessary to examine each 
individual case in detail; and N33 provides a rare opportunity 
for this research work.  \\

This paper is organized as follows. Section 2 presents the observations, 
data reduction, and the archive data used ({Spitzer data, 2MASS data). 
Section 3 describes our results (overall view, physical parameters, 
extinction, cluster search, stellar content and the field population, N33 
as a HEB, and Spitzer SED fitting). 
Section 4 presents our discussion, and finally our conclusions are summarized in Sect. 5.

\section{Observations and data reduction}

\subsection{NTT Imaging}

N33 was observed on 26 and 27 September 2002 using the
ESO New Technology Telescope (NTT) equipped with the active optics and
the Superb Seeing Imager \citep[SuSI2;][]{D'Odorico98}. 
The detector consisted of two CCD
chips, identified as ESO \#45 and \#46.  The two resulting frames were 
automatically combined to produce a single FITS file, whereas the space
between the two chips was ``filled'' with some overscan columns so that
the respective geometry of the two chips was approximately
preserved. The gap between the chips corresponds to \ab\,100 true CCD
pixels, or \ab\,8\frac.  The file format was 4288\,\x\,4096
pixels, and the measured pixel size  0\frac.085 on the sky. Each
chip of the mosaic covered a field of 5\min.5\,\x\, 2\min.7. We refer to the
ESO manual SuSI2 for more technical information.  \\

Nebular imaging was carried out using the narrow-band filters centered
on the emission lines \ha\, (ESO \#884), \hb\, (\#881), and
\oiii\,(\#882) with basic exposures of 300 sec (two exposures each on 26 September,  
six exposures for \ha, and four exposures for \hb\ and \oiii\, respectively on 27 September). 
The image quality was quite good during the night, being represented 
by a seeing of 0\frac.6.   
We constructed the line-ratio maps \ha/\hb\, and  \oiii/\hb\  
from nebular imaging. We also took exposures using filters ESO
\#811 ($B$), \#812 ($V$), and \#813 ($R$) with unit 
exposure times of 15 sec for $B$ and $V$ and 10 sec for $R$,  
respectively. The exposures for each filter were repeated twice  
using ditherings of 5\frac\,--10\frac\, for bad pixel 
rejection. \\

PSF-fitting photometry was obtained for all filters using the DAOPHOT package under 
IRAF\footnote{http://iraf.noao.fr}. The magnitudes were then calibrated using the photometric 
calibration package photcal. To perform this calibration, seven standard stars, 
belonging to two Landolt photometric groups SA\,92 and T\,Phe 
\citep[][]{Landolt92} were observed at 
four different airmasses. This led to the determination of the photometry coefficients 
and zero-points. Those coefficients are in good agreement with the indicative 
values displayed on the SuSI2 web page.\\

The aperture corrections were calculated as 
follows. Starting from one of the frames, we subtracted  
all stars except the ones used for determining the PSF 
with the daophot.substar procedure, using our preliminary 
DAOPHOT photometry and the corresponding PSF. This leads 
to a frame with only a few bright, isolated stars plus residues 
from the subtraction. We then performed both aperture and 
PSF-fitting photometry on those stars, using the same aperture 
we used for standard stars. The comparison led to aperture corrections  
of 0.02, 0.04, and 0.03 mag in $B$, $V$, and $R$, respectively.\\

During the photometry process, some slight discrepancies 
between the intensity of the frames were found: this effect was 
considered to be the consequence of episodic variations in the sky 
transparency by 7$\%$ at most. In order not to introduce a systematic 
underestimation of star magnitudes when averaging the frames, we decided to 
perform photometry on each individual frame.\\

By cross-correlating the positions of the sources in the various photometry 
files, we obtained the mean magnitude (average of the 2 mag of each filter) and a 
decent estimator of the uncertainty in this magnitude (difference between maximum 
and minimum magnitudes). Finally, the process yielded the 
photometry of 2400 stars in all three filters. The results for the brightest stars 
towards N33 are presented in Table\,\ref{tab:stars}. The whole photometry is 
available in electronic form.

\subsection{NTT spectroscopy}

The EMMI spectrograph \citep[][]{Dekker86}  
attached to the ESO NTT telescope was used on 29  
September 2002 to obtain several long-slit stellar spectra. 
The grating was \#\,12 centered on 4350\,\AA\, (BLMRD mode)
and the detector was a Tektronix CCD TK1034 with 1024$^{2}$ pixels of
size 24 $\mu$m.  The covered wavelength range was 3810-4740\,\AA\, 
and the dispersion 38\,\AA\,mm$^{-1}$, giving {\sc fwhm} 
resolutions of $2.70\pm0.10$ pixels or $2.48\pm0.13$\,\AA\, for a 1\frac.0 slit. 
At each position, we took three 10 min exposure. The instrument response was derived 
from observations of the calibration stars  LTT\,7379, LTT\,6248, and  LTT\,7987. 
The seeing condition was 0\frac.8 ({\sc fwhm}). The identifications of the stars 
along the slits (see Fig.\,\ref{fig:carte-champ} for the orientations) 
were based on monitor sketches drawn during the observations.  \\

Furthermore, EMMI was used on 26 September 2002 to obtain nebular 
spectra with gratings \#\,8 (4550-6650\,\AA) and \#\,13 4200-8000) in the REMD mode 
and with grating \#\,4 (3650-5350\,\AA) in the BLMD mode. In the REMD mode,    
the detector was  CCD \#\,63, 
MIT/LL, 2048\,\x\,4096 pixels of  15$^{2}\,\mu$m$^{2}$ each.   
Spectra were obtained with the slit set in 
east-west and north-south orientations using a basic exposure time 
of 300 sec repeated several times. The seeing conditions varied between 
0\frac.7. 
Reduction and extraction of spectra were performed using the IRAF software
package. Fluxes were derived from the extracted spectra with the 
IRAF task SPLOT. The line fluxes were measured by fitting 
Gaussian profiles to the lines as well as by simple pixel integration in 
some cases. The nebular line intensities were corrected for interstellar 
reddening using the formulae given by \citet{Howarth83} for the LMC extinction,  
which is very similar to that of the SMC in the visible. 
The intensities of the main nebular lines
are presented in Table \ref{tab:flux} where $F(\lambda)$ and $I(\lambda)$ represent
observed and dereddened line intensities. The uncertainties are
indicated by the capital letters : A $<$\,10\%, B=10--20\%, C=20--30\%, and D
$>$\,30\%. \\

\subsection{Archive Spitzer and 2MASS data}

The Spitzer archive data used in this paper come from the S$^{3}$MC
project. This is a project to map the star-forming body of the SMC
with Spitzer in all seven Infrared Array Camera (IRAC) 
and Multiband Imaging Photometer for Spitzer (MIPS) bands. 
We used the IRAC data, obtained  in 2005 May, to build a composite image of 
N33 and also obtain photometry. The typical PSF of the IRAC images in the 
3.6, 4.5, 5.8, and 8.0 $\mu$m bands is 1\frac.66 to  1\frac.98  
\citep[][]{Bolatto07}. The derived photometry for N33 in the 3.6, 4.5, 5.8, and 8.0 $\mu$m 
bands are 11.64, 11.20, 9.61, and 7.91 mag, respectively, using an integration 
aperture of 3 pixels, or 3.6\frac\, in radius 
\citep{Charmandaris08}. Measurements with either slightly larger 
or smaller apertures do not affect the color results.  
As for the MIPS fluxes (magnitudes) at the 24 and 70 $\mu$m bands,  
they are 0.13 Jy (4.37) and 1.28 Jy (-0.54) respectively.  
To examine a large field around N33 
(\ab\,400\,\x\,400 pc), we  used the SAGE-SMC observations.  
SAGE-SMC is a Spitzer Legacy program that has mapped the
entire SMC with IRAC and MIPS. The full mosaics are available at the 
Spitzer Science Center homepage, SAGE-SMC Data Deliveries. \\

We also used the {\it JHK} photometry provided by the 2MASS point source catalog 
(http://tdc-www.harvard.edu/catalogs/tmpsc.html), as presented in  
Table\,\ref{tab:stars}. Note that the embedded stars in the \hii\ region 
(\#1, \#2, and \#3) are not resolved in 2MASS data so the {\it JHK} photometry of N33-1 
corresponds to the whole N33 compact \hii\ region.

\section{Results}

\subsection{Overall view}

The images taken with the NTT telescope (Sect. 2.1) have a 
whole area of \ab\,5\min\,\x\,5\min\, corresponding to 
\ab\,90 pc\,\x\,90 pc for a distance of \ab\,60 kpc, or  {\it m\,-\,M} = 
18.94 mag \citep[][]{Laney94}. They show a starry field marked by very faint diffuse 
nebulosity running in the area. 
A close-up is presented in Fig. \ref{fig:ntt-small}, 
which displays the compact \hii\  region N33 towards the center of the image. It has a 
mean angular radius, ($\theta_{\alpha}.\theta_{\delta})^{1/2}$, of 3\frac.7 corresponding 
to a radius of 1.1 pc. Broad-band images in $B$, $V$, and $R$ (Fig.\,\ref{fig:carte-champ}) 
show the presence of three stars towards the \hii\ region (\#1, \#2, and \#3),   
whose positions and photometry are listed in Table\,\ref{tab:stars}. 
The angular separations between these stars are 2\frac.5 (\#1 and \#2), 2\frac.5 
(\#1 and \#3), and 3\frac.5 (\#2 and \#3).
We  show that the central star \#1 is the exciting source of the \hii\ 
region and therefore physically associated with the nebula. As for stars \#2 and 
\#3, they may be linked to the \hii\ region or alternatively just random field stars. 
We refer to Section 4 for a discussion. Table 1 also contains results for the brightest stars of 
the field close to the \hii\ region. Star \#6 (also known as SMC 013740 and 
2MASS J00493037-7326501) is an SMC supergiant 
K3\,I \citep[][and below Sect. 3.4]{Levesque06}. 
As for star \#4, i.e. the brightest in the vicinity of N33, it is a 
Galactic dwarf G5\,V  (see Sect. 3.4). 
The cross indicates the position of the candidate planetary 
nebula \citep[][]{Jacoby02}. Although these authors 
measure a diameter of 7\frac\, for the planetary nebula, i.e. similar to N33's, 
this object does not appear in our image. The Spitzer image, a composition 
of 4.5, 5.8, and 8.0 $\mu$m bands (Fig.\,\ref{fig:spitzer}),  
shows N33 as a red nebular object with an interesting diffuse 
arc or plume hovering over its eastern and north-eastern side. 
The nature of this feature is not yet clear. \\

An interesting aspect of the \hii\ region N33 is its isolated character, 
which we deal with amply in Sects. 3.4 and 4,. As a matter of fact, 
no star clusters and emission nebulae are visible in the whole 
\ab\,90\,\x\,90 pc field of the NTT images.  We note however that at the position of  
N33, \cite{Bica95} identify a compact object, which they classify as an emission 
nebula, situated {\sl in} the nebula DEM44 \citep{DEM76}.  
Nevertheless, a positional comparison of N33 with DEM44 
(located at 0:48:58.53 $-$73:25:39.76, J2000) shows
that N33 lies about 130~pc away from the center
of the nebula, and about 70~pc from its closest edge,
considering its dimensions of 7\arcmin~$\times$~5\arcmin,
as provided by \cite{DEM76}. As a consequence, there is no
apparent association of N33 with DEM44. 
We also examined a larger field, \ab\,400\,\x\,400 pc, 
around N33 using 
the Spitzer SAGE-SMC archive data. Similarly, no conspicuous emission source is present 
out to a projected distance of \ab\,200 pc north-west of N33. At that position, 
an extended emission object shows up, that should correspond to 
the SMC \hii\ regions N22/N23. Moreover, a minor source is seen south of N22/N23 at a projected 
distance of \ab\,160 pc from N33.

\subsection{Physical parameters}

The total H$\beta$ flux of the compact \hii\ region N33 was derived using the following 
procedure. First we calculated the relative H$\beta$ flux in an 
imaginary 1\frac\, slit passing through the H$\beta$ image with respect to 
the total flux emitted by the whole \hii\ region. This value was then 
compared with the absolute flux obtained from the spectra. The total H$\beta$ flux thus obtained was
$F$(\hb )\,=\,1.05\,\x\,10$^{-12}$ erg cm$^{-2}$ s$^{-1}$. 
Studies of the extinction in the LMC and the SMC reveal 
reddening laws that are similar to the average Galactic law for the optical 
and near-IR regions \citep[][]{Howarth83,Prevot84,Bouchet85}.
Considering the extinction law for the LMC \citep{Howarth83}, we computed the 
reddening corrected intensity $I$(\hb )\,=\,4.91\,\x\,10$^{-12}$ erg cm$^{-2}$ s$^{-1}$. 
We derived a luminosity of 2.2\,\x\,10$^{36}$ erg s$^{-1}$, or 550 \slum, for 
N33 at H($\beta$). 
This luminosity corresponds to a flux of 5.4\,\x\,10$^{47}$ \hb\  photons s$^{-1}$, or  
a Lyman continuum flux of 4.7\,\x\,10$^{48}$ 
photons s$^{-1}$ for the star, assuming that the \hii\ region is ionization bounded. 
The exciting star needed to provide this flux should have an effective temperature of 
\ab\,36,000 K or be of spectral type about O6.5-O7\,V, 
for Galactic metallicity \citep[][]{Martins05}. 
This may however be a lower limit because of photon loss in a density-bounded \hii\ region. \\

A number of the derived physical parameters of the compact \hii\ region are summarized in 
Table\,\ref{tab:param}. The mean angular radius of the \hii\ region, corresponding to 
the FWHM of cross-cuts through the \ha\ image, is given in Col. 1. The corresponding 
physical radius, obtained using a distance modulus of 
{\it m\,-\,M} = 18.94 mag \citep[][]{Laney94} is presented in Col. 2. The dereddened 
\hb\ flux obtained from a reddening coefficient of {\it c}(\hb\,) = 0.67 is given in Col. 3. 
This reddening coefficient, derived from the mean  \ha\,/\,\hb\ ratio of 4.6, 
corresponds to the whole \hii\ region. It is different from the value 
found from the nebular spectrum (Table\,\ref{tab:flux})  
because, in contrast, the spectrum belongs to a particular position and therefore 
does not cover the whole region. 
The electron temperature calculated from the forbidden-line ratio 
\oiii\,\lam \lam\,4363/(4959 + 5007), with an uncertainty 
of 4\%, is given in Col. 4. 
The electron density, estimated from the ratio of the \sii\ doublet  \lam \lam\,6717/6731, 
 is presented in Col. 5. It is accurate to \ab\,80\%.  
It is well-known that the  \sii\ lines characterize the low-density 
peripheral zones of \hii\ regions (see below for corroboration). 
Column 6 gives the rms electron density, {\it $<$n$_{e}$$>$}, 
calculated from the \hb\ flux, the radius, and 
the electron temperature, {\it T$_{e}$}, assuming that the \hii\ region is 
an ionization-bounded Str\"omgren sphere. Furthermore,  
the total mass of the ionized gas, calculated from the {\it $<$n$_{e}$$>$} 
with the previously noted Str\"omgren sphere assumption is presented in Col. 7. The ionization 
is produced by Lyman continuum photon flux given in Col. 8.

\subsection{Extinction}

The average value of the Balmer decrement towards the \hii\ region N33 is 
about 4.6, corresponding to A$_{V}$ = 1.5 mag, using the extinction law 
for the LMC with $R$ = 3.1 \citep{Howarth83}.
The most extincted part of the \hii\ region is its northeast 
border, where the \ha\,/\,\hb\ ratio reaches a value of 6.5 (A$_{V}$ = 2.5 mag). 
Interestingly, the IR elongated structure (Sect. 3.1), 
which runs from north    
to south-east (i.e. from the area indicated by 
the upper circle in Fig.\,\ref{fig:spitzer}), happens to be adjacent to 
this higher extinction area. The extinction towards star \#1 can be derived from a 
second method. O-type stars have an intrinsic color of {\it B\,--\,V}\,=\,-0.28 mag 
\citep[][]{Martins06}. This yields a color excess of {\it E(B\,--\,V)}\,=\,0.64 mag 
or a visual extinction of A$_{V}$ = 2.0 mag for star 
\#1, in good agreement with the result from the Balmer decrement. \\

Moreover, the extinction towards N33 was estimated  
by a third method using radio continuum observations. N33 appears as the 
source B0047-7343 in the Parkes
radio continuum survey at 3 and 6 cm, which had beam-sizes of 
2\min.7 and 13\min.8, respectively \citep[][]{Filipovic98}. 
Higher resolution observations of this object in the same 
wavelength range were obtained by \citet[][]{Indebetouw04}, 
who in their search for ultracompact and buried \hii\ regions  
in the Magellanic Clouds, used the Australia Telescope Compact  
Array (ATCA) in radio continuum emission at 3 and 6 cm with 
synthesized beams of 1\frac.5 and 2\frac, respectively. 
Since the beam-widths are 1\frac.5 and 2\frac, respectively, the radio continuum 
observations do not sample the entire \hii\ region. To compare the radio 
continuum fluxes with that of \hb , we corrected the \hb\ flux. 
We computed these from the ratio of 
the measured \hb\ emission flux with different apertures (1\frac.5 and 2\frac\ ) 
to the total N33 flux.
 The resulting extinction, A$_{V}$ = 1.9 mag, although comparable with that obtained 
using the previously mentioned methods in the optical range, 
may be an underestimate 
if the flux is not uniformly distributed over the \hii\ region and the smaller radio 
lobes miss a part of it.

\subsection{Cluster search}

We perform a cluster analysis technique on the stellar 
photometric catalog from our SuSI2 imaging, to identify any 
potentially important stellar concentration at the surroundings of the 
compact {\sc H~ii} region N33 that would suggest membership of its 
central star to a stellar cluster. Our identification method is based 
on star counts in quadrilateral grids for the construction of stellar 
density maps in the area of interest \citep[see e.g.,][]{Gouliermis00}.  
We performed star counts on our complete photometric catalog 
of 2400 sources. This was done by counting the stars in grids according to their 
celestial positions, as they are defined by our astrometry.  The method is 
sensitive to three factors, namely 1) the stellar numbers, 2) the size 
of each grid element, which defines the smallest possibly detectable 
cluster (the resolution element in the density maps), and 3) the size 
of the available field-of-view (FoV). This is due to the isopleths of the 
stellar density maps whose steps are defined to be equal to the standard deviation, $\sigma$, 
of the {\sl total} average stellar density (of stars per grid element) of the 
considered field. We refer to the latter as the ``background density''. We 
consider any stellar concentration revealed by the isopleths, that 
correspond to stellar density of 3$\sigma$ above the background and 
higher, as {\sl statistically important}, and therefore as 
a candidate true cluster. \\

The application of star counts to the whole observed FoV  
allows us to identify any possible membership of N33 to any large 
stellar concentration of size on the order of $\sim$~100~pc. It should be noted that the
gap between the chips of SuSI2 introduces a lack of stars along it in
every photometric catalog derived for each filter. As a consequence, the
final combined photometric catalog also contains a lack of stars along a
100 pixel thick line vertically crossing the middle of SuSI2 frame. 
While this introduces a problem into the consistency of the detection, the gap is 
fortunately away from the \hii\ region, and hence its effect 
on the cluster detection at the immediate vicinity of N33 is not 
important. We constructed the stellar density maps for the SuSI2 FoV 
using a grid of 25 $\times$ 25 elements, corresponding to a physical 
scale of about 3~pc per grid element, which defines the smallest 
possible stellar concentration that can be revealed in the isodensity 
maps. These maps are constructed for all the stars, as well as for 
selected ``blue stars'' with $B-V \leq 0.55$ mag, mostly 
representing the main-sequence stellar population and `red stars'  
selected to have colors $B-V > 0.55$ (see color-magnitude diagram of Fig.\,\ref{fig:hr}), 
which represent the evolved stellar population of the general SMC field. In 
Fig.~\ref{fig:contours}, the stellar isodensity maps at the vicinity of 
N33, i.e.  in the close-up field of Figs.~\ref{fig:ntt-small} and  \ref{fig:carte-champ}, 
are shown along with the corresponding stellar map. \\

As seen from the density maps of Fig.~\ref{fig:contours}, all 
the isopleths around N33 appear below the 3$\sigma$ density 
threshold. This clearly implies that, considering the factors on which 
the star counts method depends, there is no statistically 
significant stellar cluster around N33, of size larger than 3 pc. 
Our search for stellar clustering around N33 is based on the 
complete stellar catalog constructed from our SuSI2 photometry. This catalog, 
according to the evolutionary models, covers stars with masses 
as low as about 2\,\sm\, corresponding to late A-type dwarfs. 
It should be noted that the low completeness in faint stars 
may not allow us to have a significant density peak around N33 simply 
because we do not see these stars. Nevertheless, a small stellar clump 
of projected size $\simeq$~18\frac\,\x\,26\frac\ 
(5~pc~$\times$~7~pc) is  revealed around N33 at the $1\sigma$ 
density level and the isodensity maps of Fig.~\ref{fig:contours} 
suggest that it mainly consists of main-sequence stars. However, its 
low density, reaching the 3$\sigma$ level only at its very central 
peak, does not classify it as a real cluster. Moreover, from our 
cluster analysis in the whole observed FoV we assess that {\sl N33 
does not seem to belong to any larger stellar concentration on a 
100~pc length-scale}. A possible reason for not identifying any parent 
stellar structure for N33 is that the \hii\ region may indeed be 
part of a larger stellar aggregate, the size of which is much larger 
than our observed FoV. In this case, our search for important density 
enhancements would take place {\sl within} the system itself, without 
apprehending its existence, thus our average background density 
would not represent the real {\sl background}, but the high stellar 
density levels of the large concentration. However, previous 
searching-by-eye investigations by \cite{Hodge85} and \cite{Bica95}, 
on FoVs larger than ours do not detect any large stellar concentration 
towards this region.  The only indication of a large stellar concentration in this
region is found by \cite{battinelli91}, who identified a 
candidate stellar association, that seems to coincide with 
the larger nebula DEM44. This author applied a cluster analysis 
technique, the {\sl path linkage criterion}, to previous catalogs 
of OB-type stars to identify candidate OB associations in the SMC. 
However, this method is applied only to the OB stars, and therefore 
is biased towards large-scale ``connections'' of these stars without 
any supporting information about their physical relation or any 
interaction to each other. Moreover, the identification of large OB 
groups based solely on the positions of the stars can be challenged, because it 
has been confirmed that young stellar systems may lose a significant 
fraction of their massive stars at the very beginning of their evolution, 
\citep[e.g.,][]{Gvaramadze08}, and therefore OB stars may be mistakenly 
considered as members of a system, while being runaways from another. 
Under these circumstances, and because DEM44 does not seem
to contain N33, the nebula coinciding with a candidate OB association 
cannot be considered proof
that N33 does belong to this large stellar concentration. \\

To assess the clustering behavior of the observed stars 
in the region of N33, we construct the two-point correlation function (TPCF), 
which determines the distance between all possible pairs of stars. This method, 
as applied by e.g. \cite{gomez93} to study the distribution of young stars 
in a Galactic star-forming region,  considers the excess number of pairs in the 
actual distribution over a random distribution. Here, we apply the method as extended 
by \cite{larson95}, who introduced the assessment of stellar clustering in 
terms of the average surface density of companions, $\Sigma_{\rm c}(\theta)$, as 
a function of the projected angular separation, $\theta$.  We measured 
$\Sigma_{\rm c}(\theta)$ and correlated it to the corresponding angular 
separation for the blue stars in the whole observed FoV, as well as for the stars 
covered by the close-up field centered on N33. The constructed TPCFs are shown 
in Fig.~\ref{fig:tpcf} drawn with a red line for the close-up field and blue for the 
whole observed field. In both samples, a clear change in the correlation 
$\Sigma_{\rm c}(\theta)$ appears at $\theta\simeq$~10\arcsec, which at the distance 
of the SMC corresponds to about 3~pc. We note that this limit 
is similar to that measured for the small stellar clump seen at 
the position of N33 in the density maps of Fig.~\ref{fig:contours}. 
For larger separations ($\theta > 10\arcsec$), the relation $\Sigma_{\rm
c}(\theta)$ becomes almost flat, with 
$\displaystyle \Sigma_{\rm c} \propto \theta^{-0.2}$. This almost-constant
density of companions for larger separations indicates a random distribution of 
stars and therefore {\sl no clustering of stars at scales larger 
than 3~pc}. On the other hand, for smaller scales the average density of 
companions correlates linearly with the separation as 
$\displaystyle \Sigma_{\rm c} \propto \theta^{-1.1}$, 
providing clear evidence of self-similar clustering on scales as small as 
the smallest resolved 
separation, and eventually in the binary and multiple systems regime. This implies 
that stellar clustering in the observed FoV seems to occur only at scales 
smaller than about 3~pc, in agreement with the findings of our cluster analysis above. 
In conclusion, from our analysis we find that the \hii\ region N33
does not seem to belong to any large stellar concentration. 
On smaller scales, we identify a peak in the stellar
density around N33, with a stellar mass of roughly 100~\sm, which
nevertheless is very loose and therefore should be considered as a random
density fluctuation rather than a real cluster.

\subsection{Stellar content and the field population}

We obtained two spectra in our program of stellar spectroscopy (Sect. 2.2). 
However in spite of the relatively good seeing conditions (0\frac.7 {\sc fwhm}), 
extracting uncontaminated spectra is not straightforward. The compact 
\hii\ region (mean size 7\frac.4) has strong emission lines that fill in the 
absorption lines of the embedded stars, in particular those of \hei . 
Nevertheless, \heii\, absorption lines at \lam \lam\,4200, 4541, and 4686 are 
certainly present, albeit with weak S/N ratio, in the spectra uncorrected for 
nebular lines (Fig.\,\ref{fig:spectres}). This agrees with the presence 
of an O6.5-O7\,V spectral type, as inferred from the \hb\ flux measurement 
(see Sect. 3.2). The estimated spectroscopic mass of an O6.5\,V star is 
29\,\sm\ and its effective temperature  $T_{\rm eff}$\,=\,36,800 K \citep[][]{Martins05}. 
We also note that  \citet[][]{Indebetouw04} derive a spectral type of O8.5\,V for the exciting 
source of N33 from their radio continuum observations at 3 and 6 cm. As underlined 
above, these ATCA observations do not sample the entire \hii\ region.  
Furthermore, the presence of the \heii\, absorption lines in the spectrum of the 
exciting star rules out a confusion of N33 with a planetary nebula.  \\

The spectrum of star \#4, the brightest object close to the \hii\ region, is 
presented in Fig.\,\ref{fig:spectres}. Using classification criteria for late-type stars 
\citep[][]{Jaschek87,Gray09}, star \#4 can be assigned the subtype G5. This spectral type 
is based on the strength of the \caii\ $\lambda$4227 line and the ratios 
\fei\ $\lambda$4046/H$\delta$ and \fei\ $\lambda$4325/H$\gamma$. 
This result agrees with the measured colors: $B - V$ = 0.78, $V - R$ = 0.39, 
$V - K$ = 1.76, $H - K$ = 0.07, and $J - K$ = 0.42 mag \citep[][]{Koornneef83}.
In terms of luminosity class, it is a dwarf, as suggested by the line ratios  
\srii\ $\lambda$4077/\fei\ $\lambda$4046 and 
\yii\ $\lambda$4376/\fei\ $\lambda$4383. We also note that the \caii\ K and H lines 
do not have extremely broad and damped wings as expected in supergiants. 
This star should have an absolute magnitude $M_{V}$ = +5.1 \citep[][]{Jaschek87}. 
However, this value is at odds with the observed absolute magnitude of $M_{V}$ = -7.31 
assuming that it belongs to the SMC. This discrepancy indicates that star \#4 is 
a Galactic member. Additional evidence of the Galactic membership of this star is its  
relatively low radial velocity. In contrast, star \#6, the other bright field star,  
has a radial velocity of 156.4 km s$^{-1}$ which confirms its SMC membership.
Figure\,\ref{fig:spectres} also displays the spectrogram of star \#6, 
which is much redder ($V - K$ = 4.17 mag, see also Table\,\ref{tab:stars}). 
We classify this star as an early-K type supergiant, in agreement with its previously 
reported classification,  K3\,I \citep[][]{Levesque06}.\\

Figure \ref{fig:hr} displays the color-magnitude diagram of the population 
of stars in the entire NTT field of 304\frac\,\,\x\,316\frac\,  
(90\,\x\,93 pc), for a cut-off magnitude of $V$ = 20.5. 
We also overplotted four isochrones with ages 18 Myr, 56 Myr, 500 Myr, 
and 1 Gyr for metallicity Z = 0.004 \citep{Lejeune01}. A fifth isochrone of 
14 Myr is calculated with a higher metallicity of Z = 0.008. The bulk of the stars 
are concentrated into two groups: an apparent main sequence centered on 
$B-V$\,\ab\,0  mag and an evolved population centered on $B-V$\,\ab\,0.8 mag. 
The presence of star \#6, which is  an SMC supergiant of type K3\,I 
(explained above), provides a practical constraint in choosing the suitable 
isochrones. Star \#6 is appropriately fitted by two isochrones, those of 
18 Myr and 14 Myr. The latter is the best fit, but assumes a different 
metallicity from the other isochrones, i.e. a factor of two higher than 
the average value for the SMC. This higher metallicity may represent the effect of 
heavy element enrichment as the K3 star evolves to higher luminosities.  
These 14 Myr and 18 Myr isochrones also closely reproduce the properties of 
the main sequence stars, which in their vast majority have an initial mass of 
roughly $<$ 8\,\sm.\ \\ 

As far as the exciting star of the \hii\ region N33 is concerned, i.e. star \#1, 
it is  indicated by a cross on the color-magnitude diagram. 
This star is affected by an extinction of A$_{V}$ = 2.0 mag, so it should not  
belong to the above-mentioned low- and intermediate-mass population, which are mostly 
spread around $B - V$\,\ab\,0 mag. Star \#1 appears to be  
a young massive star occurring in a field of 
old low-mass stars. We also note the color-magnitude diagram position  
of star \#2, which apparently lies across 
the \hii\ region N33. If this star really belongs to the 56 Myr isochrone, as 
suggested by the diagram, this rules out its association with the \hii\ region. 
As for star \#3, it seems to be consistent with the position of a reddened 
B-type star, hence its association with the \hii\ region cannot be excluded 
at this stage. 
The second group of stars on the color-magnitude diagram, 
whose most evolved members gather between $B-V$ colors  
0.5 and 1.5 mag, have a possible age ranging from 500 Myr to 1 Gyr   
and a turn-off of \ab\,2\,\sm. A large number of stars belonging to these 
older isochrones apparently coincide with the lower part of the 
18 Myr and 14 Myr main sequences. We suggest that these various 
stellar populations may not be locally related to star \#1.     \\ 

The reason is that the SMC is known to have a complex structure with considerable 
line-of-sight depth, as shown by several works; we refer to 
\citet[][]{Westerlund90} for a review,  as well as  
\citet[][]{McGee81,McGee82}, 
\citet[][]{Staveley-Smith97}, and 
\citet[][]{Hatzidimitriou05}. In particular,  
\citet[][]{Mathewson86,Mathewson88}  
measured the distances of 161 Cepheids in the SMC, using the period-luminosity 
relation in the infrared. The Cepheids were found to extend from 43 to 75 kpc with a maximum 
concentration at 59 kpc. Since N33 is not part of an OB association 
(see above Sects. 1 and 4) and more importantly it is an isolated \hii\ region, 
the various stellar populations present in the color-magnitude diagram should not be 
physically related to it. They are probably distributed  along the line of sight over 
32 kpc. We suggest that the \hii\ region N33 may not be physically associated with this 
main sequence.

\subsection{Chemical abundances}

An example of the nebular spectra of N33 is shown in Fig.\,\ref{fig:neb}, whereas 
Table\,\ref{tab:flux} lists the corresponding main lines. The spectrum represents an \hii\ region, 
with no \heii\ lines. The two faint lines redward of \hei\ \lam\,5876 are 
\lam\,5888 and \lam\,5890 in emission. Both exist in the Orion \hii\ region, the first one  
is unidentified and the second  is a \cii\ line \citep[][]{Baldwin00}. 
Similarly, these two lines are present in the spectrum of the SMC compact \hii\ region N66A. 
The ionic abundances with respect to H$^{+}$  were calculated from nebular lines 
using the IRAF task ionic of the package NEBULAR \citep{Shaw95}. The abundance values are 
listed  in Table\,\ref{tab:ion}. \\

To derive the total abundances of a given element, it is necessary to estimate the amount of 
the  element in ionization states not observed in our spectra.
We therefore used a set of ionization correction factors (ICFs) to convert into 
elemental abundances. 
The absence of the \heii\ line indicates that He$^{2+}$/H$^{+}$ is negligible. 
Moreover, we assume that neutral helium is not important. Thus we 
assumed that the total He/H ratio is just equal to He$^{+}$/H$^{+}$.
The total abundance of oxygen was adopted to be the sum of O$^{+}$ and O$^{2+}$ abundances.
The absence of \heii\ recombination lines in our spectra and the similarity between 
the ionization potentials of He$^+$ and O$^{++}$ implies that the contribution of O$^{3+}$ 
is not significant.
To obtain the total abundance of nitrogen, we used the usual ICF
based on the similarity between the ionization potential of 
N$^{+}$ and O$^{+}$ \citep{Peimbert69}. The N$^{+}$ abundance does not depend strongly 
on the electron temperature. The largest errors come from the uncertainty in the 
\lam \lam\,6548 and 6584 line intensities. Our N result is accurate to within about 
30\%. As for Ne, the only measurable lines in the optical range are those of 
Ne$^{2+}$ but the amount of Ne$^{+}$ may be important in the region. 
We adopted the usual expression of the ionization correction factor of Ne  
that assumes that the ionization structure of Ne is similar to that of O \citep{Peimbert69}. \\

The total chemical abundances for N33 are presented in Table\,\ref{tab:ab}.
The most accurately estimated abundances belong to He and O, which are accurate to within 
15 and 20\% respectively. Table\,\ref{tab:ab} also presents the mean abundance 
values derived for the SMC  \citep[][]{Russell92}. The N33 abundances agree with 
the SMC mean values.  In other words, N33 is, as expected, a low-metallicity ionized nebula. \\

\begin{table*}             
\caption{Positions and photometry of the main stars in the field of SMC N33$^{\dag}$}  
\label{tab:stars} 
\begin{tabular}{l c c c c c c c r c}     
\hline\hline
star ID & $\alpha$ (2000) & $\delta$ (J2000) & $V$ & $B-V$ & $V-R$ & $J$ & $H$ & $K$ 
& spectral type\\
\hline
N33-1  & 00:49:29.25 & -73:26:33.99 & 15.92 &  0.36 &  0.18 & 13.94 & 13.31 & 12.96 & O6.5-O7\,V \\
N33-2  & 00:49:29.62 & -73:26:35.70 & 16.39 &  1.07 &  0.61 &       &       &       & \\
N33-3  & 00:49:28.84 & -73:26:35.28 & 17.50 &  0.28 &  0.10 &       &       &       & \\
N33-4  & 00:49:24.60 & -73:26:45.51 & 11.65 &  0.78 &  0.39 & 10.31 &  9.96 &  9.89 & 
    G5\,V (Galactic) \\
N33-5  & 00:49:41.75 & -73:26:41.83 & 13.04 &  0.77 &  0.40 & 11.72 & 11.38 & 11.33 & \\
N33-6$^{\ddag}$  & 00:49:30.58 & -73:26:50.87 & 13.56 &  1.76 &  0.95 & 10.48 &  9.62 &  9.39 & 
    K3\,I \\
N33-7  & 00:49:36.86 & -73:27:54.49 & 13.89 &  0.59 &  0.36 & 12.55 & 12.19 & 12.09 & \\
N33-8  & 00:49:14.22 & -73:26:15.80 & 14.71 &  1.48 &  0.33 & 12.10 & 11.33 & 11.18 & \\
N33-9  & 00:49:23.10 & -73:25:53.87 & 15.01 &  0.36 &  0.20 & 14.13 & 13.86 & 13.82 & \\
N33-10 & 00:49:43.17 & -73:26:37.02 & 15.03 &  0.18 &  0.11 & 14.47 & 14.30 & 14.28 & \\
N33-11 & 00:49:18.73 & -73:27:31.13 & 15.23 &  1.27 &  0.66 & 12.96 & 12.32 & 12.21 & \\
N33-12 & 00:49:39.42 & -73:27:17.54 & 15.27 & -0.19 & -0.13 & 15.75 & 15.66 & 15.45 & \\
N33-13 & 00:49:39.61 & -73:26:13.45 & 16.10 &  1.37 &  0.72 & 13.67 & 12.93 & 12.83 & \\
\hline   
\end{tabular} \\

$\dag$
The {\it BVR} photometry results from NTT observations while the 
{\it JHK} measures come from the 2MASS catalog. \\
$\ddag$
Also named SMC 013740.
\end{table*}

\begin{table*}
\caption{Some physical parameters of the compact \hii\ region SMC N33}
\label{tab:param}  
\begin{tabular}{c c c c c c c c} 
\hline\hline
$\theta$ & $r$  & $I$(\hb )$^{\dag}$        & {\it Te}    & {\it Ne}$^{\ddag}$   & {\it $<$n$_{e}$$>$}  
& {\it M$_{gas}$}  & $N_{L}$ \\
(\frac\,) & (pc) & erg s$^{-1}$ cm$^{-2}$   & (K) & cm$^{-3}$ & cm$^{-3}$ & (\sm ) & ph s$^{-1}$ \\
         &      & \x\,10$^{-12}$ &      &      &      &        &  \x\,10$^{48}$ \\
\hline
    3.7    &    1.1    &     4.91        &  12540    & 450     &  380    &   64  &  4.7 \\
\hline
\end{tabular} \\

$\dag$ Corrected for reddening with {\it c}(\hb\,)\,=\,0.67.\\
$\ddag$ Estimated from the \sii\ ratio.
\end{table*}

\begin{table*}
\caption{Nebular line intensities of the SMC compact \hii\ region N33}
\label{tab:flux}
\begin{tabular}{llccc}
\hline  \hline	
  \multicolumn{2}{c}{} &  \multicolumn{2}{c}{} &\\
		$\lambda$ (\AA) & Iden. & $F$(\lam\,) & $I$(\lam\,) 
                 & Accuracy\\
  \hline	
	3727,29 & \oii\           & 77.1  &  108.8 & A \\
	3797    & H10	          &  2.4  &   3.3 & B \\
	3835    & H9              &  3.0  &   4.1 & B \\
	3869    & \neiii\         & 15.2  &  20.3 & A \\
	3889,90 & \hei\ + H8      &  8.6  &  11.4 & B \\
	3968,70 & \neiii\ + \he\  & 11.8  &  15.2 & A \\
	4101    & \hd\            & 14.0  &  17.2 & A \\
	4340    & \hg\            & 29.1  &  33.3 & A \\
	4363    & \oiii\          &  4.8  &   5.5 & B \\
	4471    & \hei\           &  3.4  &   3.7 & C \\
	4861    & \hb\            & 100.0 & 100.0 & A \\
	4959    & \oiii\          & 148.3 & 145.1 & A \\
	5007    & \oiii\          & 444.8 & 430.8 & A \\
	5876    & \hei\           &  13.8 &  11.5 & B \\
	6300    & \oi\            &   4.4 &   3.4 & C \\
	6312    & \siii\          &   2.3 &   1.8 & C \\
	6363    & \oi\            &   1.3 &   1.0 & D \\
	6548    & \nii\,          &   4.1 &   3.1 & C \\
	6563    & \ha\            & 380.0 & 286.0 & A \\
	6584    & \nii\           &  10.3 &   7.8 & B \\
	6678    & \hei\           &   4.1 &   3.1 & B \\
	6716    & \sii\           &  10.1 &   7.5 & B \\
	6731    & \sii\           &   9.3 &   6.9 & B \\
	7065    & \hei\           &   4.3 &   3.1 & C \\
	7135    & \ariii          &  11.9 &   8.4 & B \\
	7236    & \ariv\          &   1.0 &   0.7 & D \\
	7323    & \oii\           &  10.1 &   7.0 & C \\
	7751    & \ariii\         &   3.7 &   2.4 & C \\
   \hline
   \multicolumn{1}{l}{c(H$\beta$) = 0.40} \\
   \hline
\end{tabular}
\end{table*}

\begin{table*}
\caption{Ionic abundances of SMC N33}
\label{tab:ion}
\begin{tabular}{lcc}
\hline\hline
Ion        & N33  \\
\hline
He$^+$/H$^+$                   & 0.087  \\
O$^+$/H$^+$ (\x\,$10^{5}$)     & 3.65\\
O$^{++}$/H$^+$ (\x\,$10^{5}$)  & 7.45\\
N$^+$/H$^+$ (\x\,$10^{6}$)     & 0.92 \\
Ne$^{++}$/H$^+$ (\x\,$10^{6}$) & 8.89\\
S$^+$/H$^+$ (\x\,$10^{7}$)     & 2.10\\
Ar$^{++}$/H$^+$ (\x\,$10^{7}$) & 4.73\\
\hline	
\end{tabular}
\end{table*}

\vspace{0.5cm}

\begin{table*}
\caption{Elemental abundances in SMC N33} 
\label{tab:ab}
\begin{tabular}{lccc}

\hline\hline
   Element & N33  & mean SMC$^{\dag}$  \\
\hline
He/H                & 0.087 & 0.081 \\
O/H (\x\,$10^{4}$)  &  1.11 &  1.07 \\
N/H (\x\,$10^{6}$)  &  2.76 &  4.27 \\
Ne/H (\x\,$10^{5}$) &  1.34 &  1.86 \\
\hline \\
$\dag$ \citet[][]{Russell92}
\end{tabular}
\end{table*}

\begin{figure*}[]
\centering
\includegraphics[width=1.0\hsize]{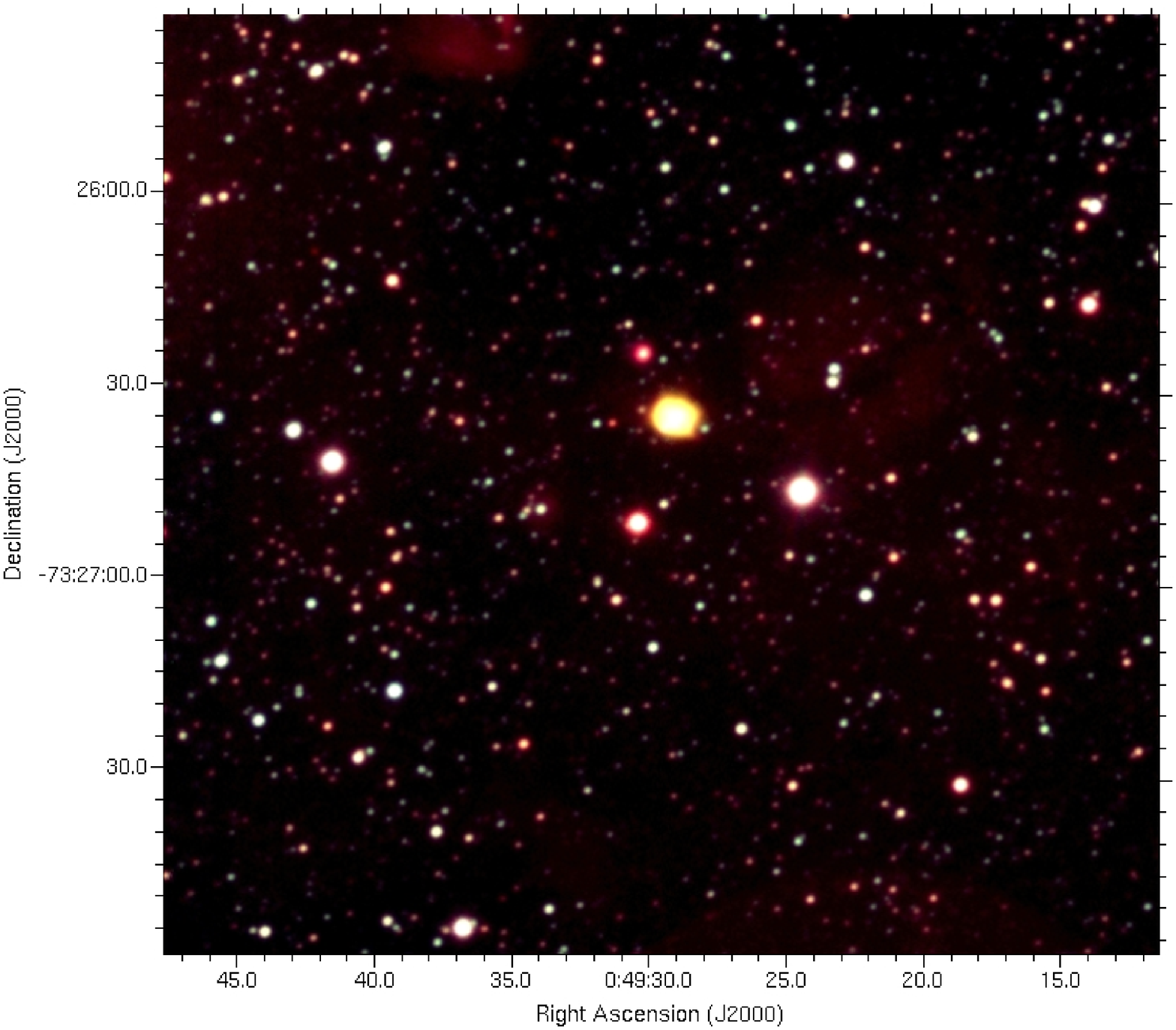}
\caption{A composite three-color image of the area of the SMC containing 
the \hii\ region N33. The \hii\ region the bright  object is situated above the image center at 
coordinates $\alpha$\,=\,00h\,49m\,29s and $\delta$\,=\,-73\deg\,\,26\min\,34\frac.  
See Fig.\,\ref{fig:carte-champ} for the identification of other objects.
The image, taken with the ESO NTT/SuSI2, results from the coaddition of narrow-band 
filters \ha\ (red),  \oiii\ (green), and \hb\ (blue). 
Field size 156\frac\ \x\,147\frac\, (\ab\ 46\,\x\,43 pc). 
This is a close up 
of an original image covering a field of  304\frac\,\,\x\,316\frac\, 
corresponding to 90\,\x\,93 pc. North is up and east to the left. }
\label{fig:ntt-small}
\end{figure*}

\begin{figure*}[]
\centering
\includegraphics[width=1.0\hsize]{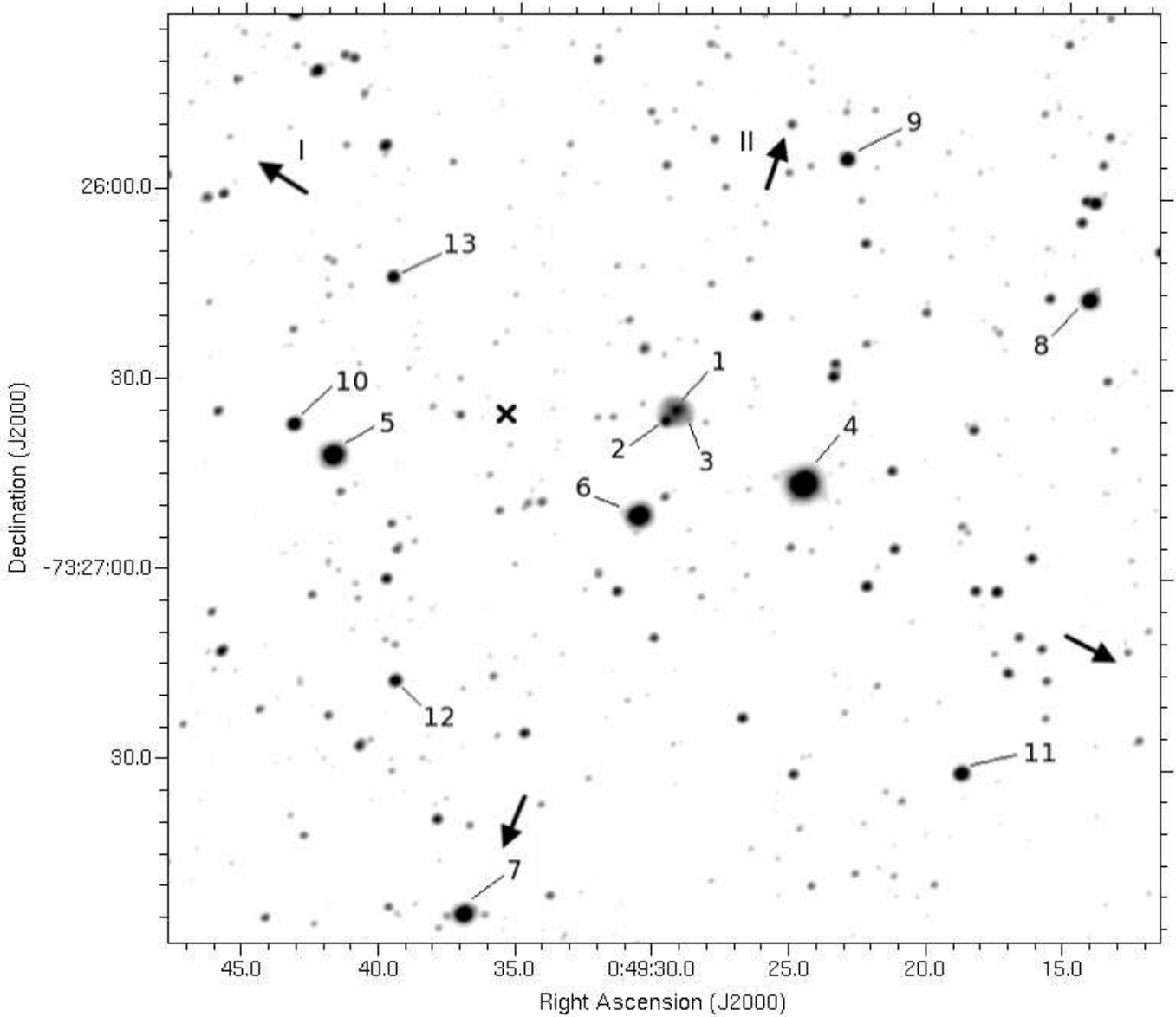}
\caption{A broad-band image of the SMC N33 region obtained through filter 
$R$ (ESO \#813). Same field size and orientation as in  Fig.\,\ref{fig:ntt-small}. 
The arrows I and II show the directions of the spectrograph slit. Seeing 0\frac.6 
FWHM. The \hii\ region apparently 
contains three stars (numbered \#1, \#2, and \#3). 
The brightest stars of the field are also numbered. 
The cross indicates the position of the candidate planetary nebula reported by 
\citet[]{Jacoby02}.
} 
\label{fig:carte-champ}
\end{figure*}

\begin{figure*}[]
\centering
\includegraphics[width=1.0\hsize]{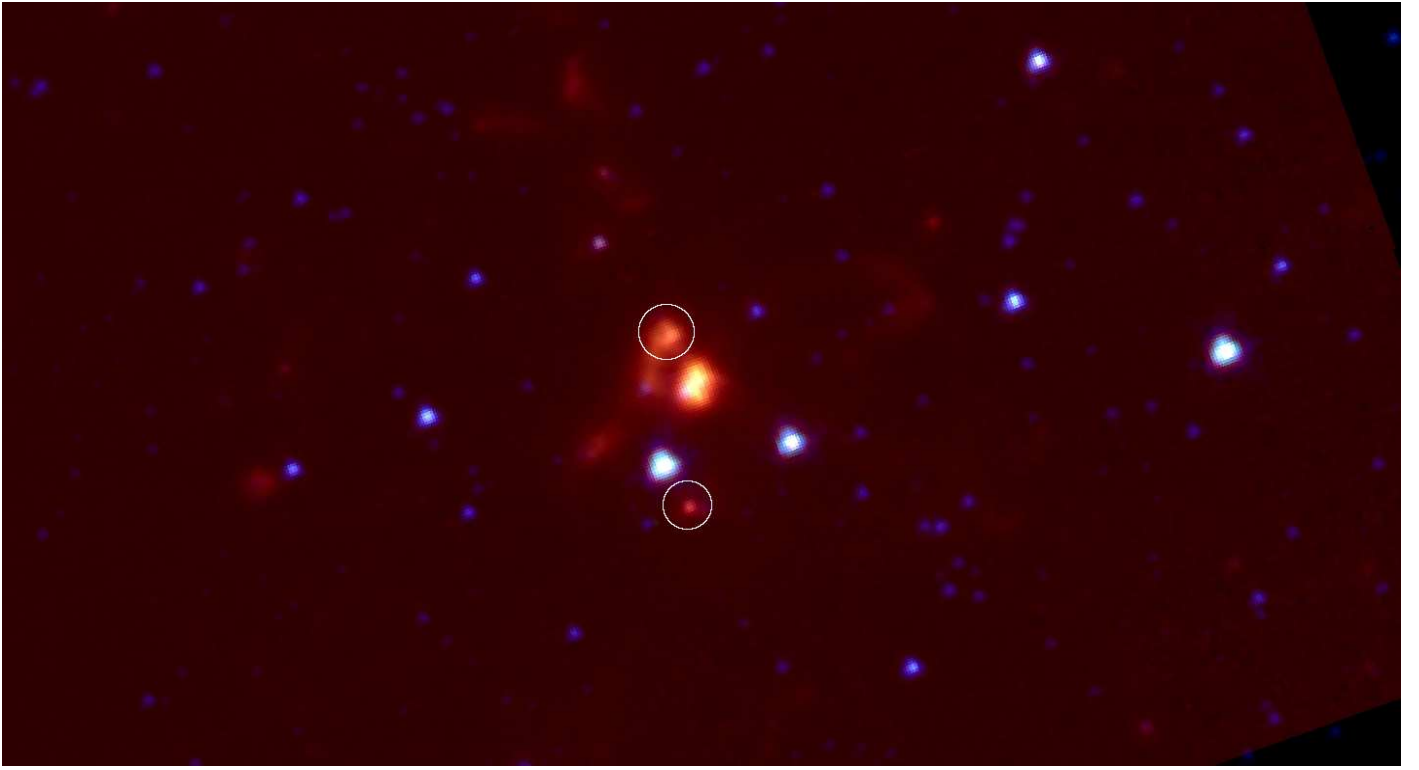}
\caption{
A composite image of the SMC N33 region obtained with Spitzer IRAC data. 
The object at the center is the \hii\ region N33. The two 
other bright sources are stars \#4 and \#6 (see Fig.\,\ref{fig:carte-champ}). 
The 4.5 $\mu$m band is represented in blue, 
the 5.8 $\mu$m band in yellow, and the 8.0 $\mu$m band in red. 
The compact \hii\ region N33 is the red nebular object at center. 
Note the diffuse emission curl over the main body of the \hii\ region running 
from north (inside the upper circle) to south-east. 
The circles enclose the young stellar object (YSO) 
candidates S3MC\_J004930.12$-$732623.42 
and S3MC\_J004929.07$-$732658.56 reported by \citet[][]{Bolatto07}. 
However, the true nature of these objects and their relation with N33 are not 
clear; see the text (Sect. 4). 
Field size 280\frac\ \x\,160\frac\, (\ab\ 83\,\x\,47 pc).  North is up and east to the left. } 
\label{fig:spitzer}
\end{figure*}

\begin{figure*}[t!]
\begin{center}
\includegraphics[width=0.975\textwidth]{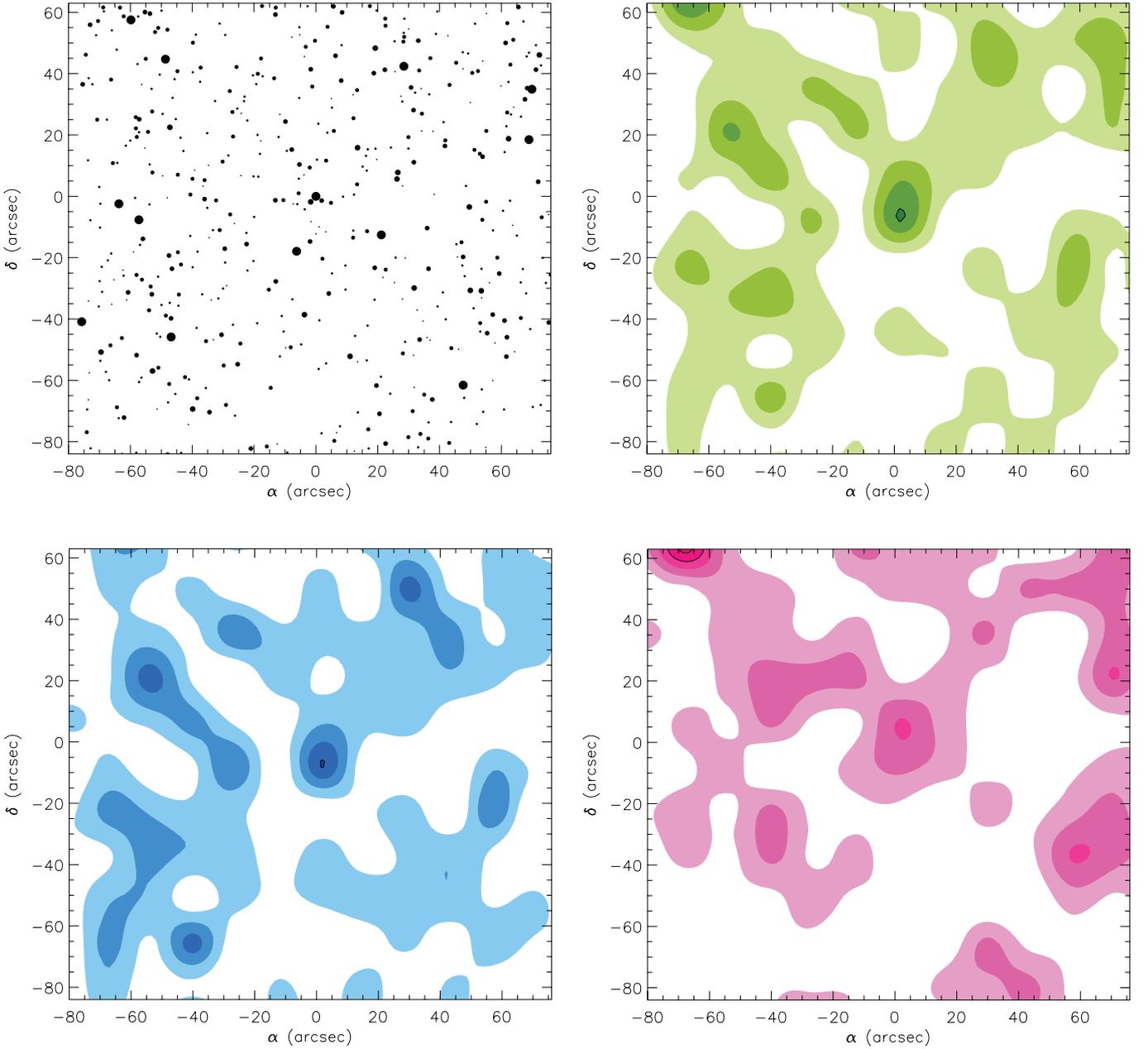}
\caption{Isodensity contour maps constructed to search for stellar clustering around the 
compact SMC \hii\ region N33. In the top-left panel, the stellar map for the close-up 
field of view centered on N33 (position 0,0). 
The stellar density maps refer to the same close-up field
of N33. They are shown  
as they are constructed for all observed stars (green map, top-right panel), for blue 
(main sequence) stars (blue map, bottom-left panel), and red evolved stars 
(red contour map, bottom-right panel). 
Isopleths are plotted starting at the average {\sl background} stellar density 
and in steps of $\sigma$ above this level. 
No stellar concentration with density $\geq 3\sigma$ above the background
is identified in the maps. Only the central part of the small stellar clump encircling N33
in the green and blue maps reaches the $3\sigma$ density level. }
\label{fig:contours}
\end{center}
\end{figure*}

\begin{figure*}[t!]
\begin{center}
\includegraphics[width=0.975\textwidth]{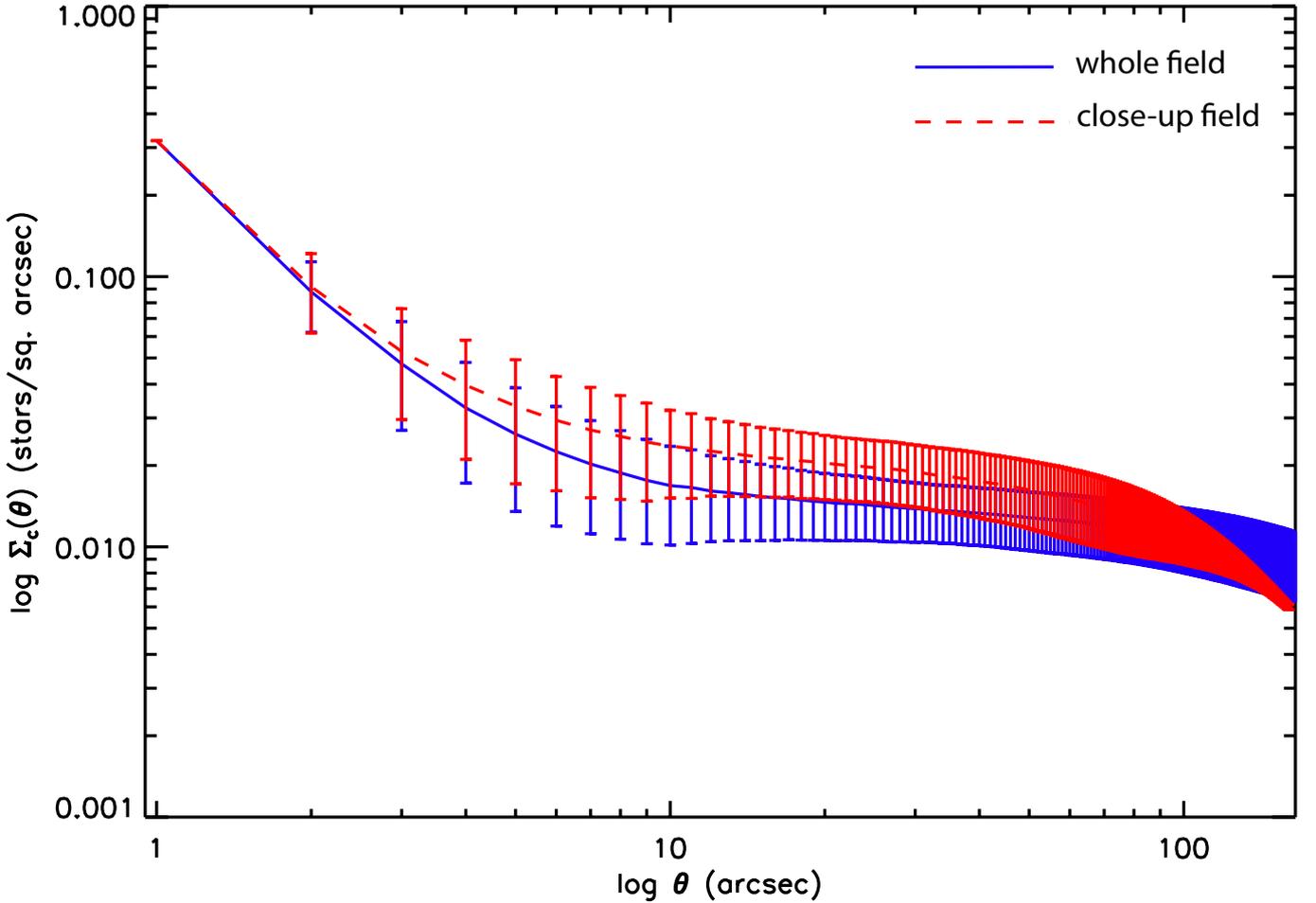}
\caption{Average surface density of stellar companions as a function of projected angular 
separation from each blue star in the whole observed area (blue solid line) 
and in the region of the vicinity of the \hii\ region N33 (red dashed line). 
Error bars are indicated. From both curves, one 
can distinguish a break in the correlation at almost the same separation of 
$\theta \simeq 10\arcsec$, i.e., 3~pc. This scale separates any large stellar 
clustering from the small binary/multiple stellar clumps.
\label{fig:tpcf}}
\end{center}
\end{figure*}

\begin{figure*}[]
  \centering
\includegraphics[width=14cm]{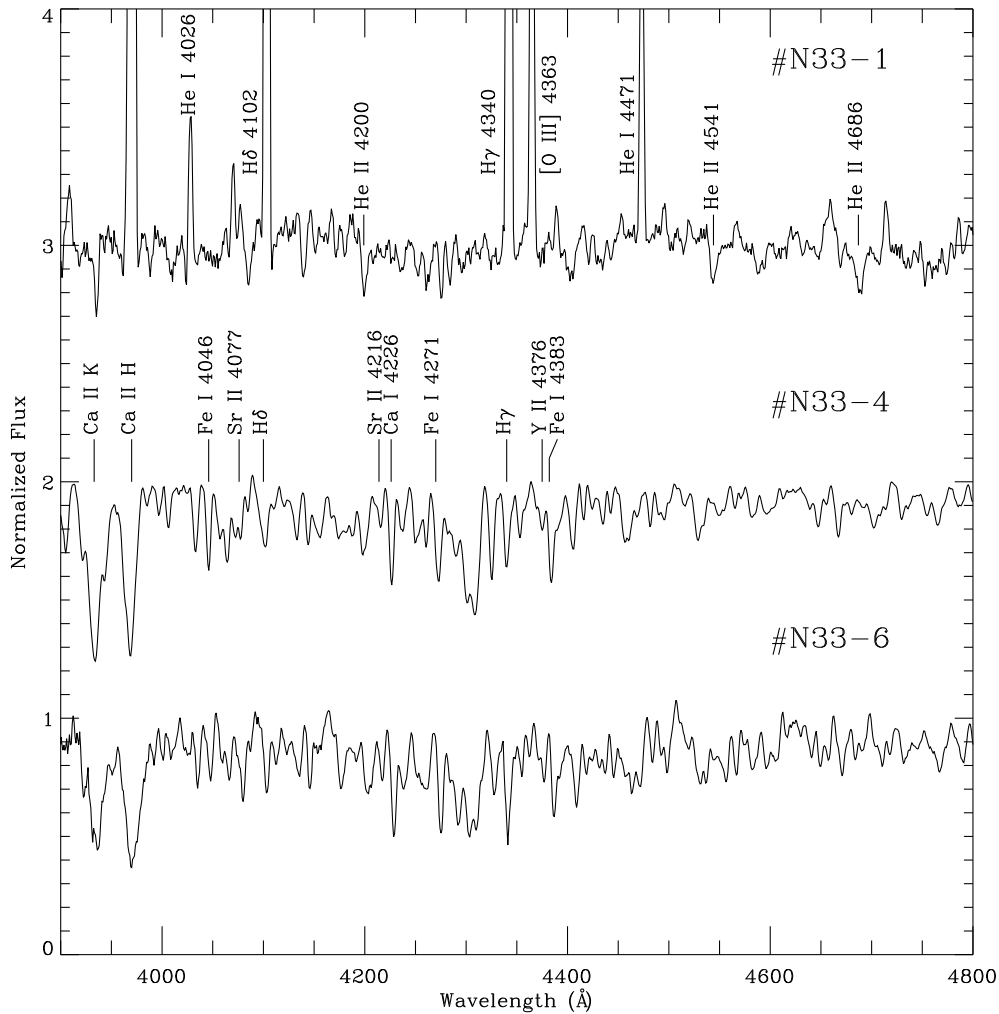}
\caption{
Spectrograms of three stars observed towards SMC N33. Star \#1 
is the exciting source of the compact \hii\ region N33. This is a raw spectrum, 
i.e. it is not corrected for nebular emission lines. Note the  
\heii\ absorption lines indicating a hot massive star (see Sect. 3.5). 
Stars \#4 and \#6, lying in the field of N33, are classified 
as G5\,V (Galactic) and K3\,I (SMC) respectively.} 
\label{fig:spectres}
\end{figure*}

\begin{figure*}[]
\centering
\includegraphics[width=13.5cm,angle=-90]{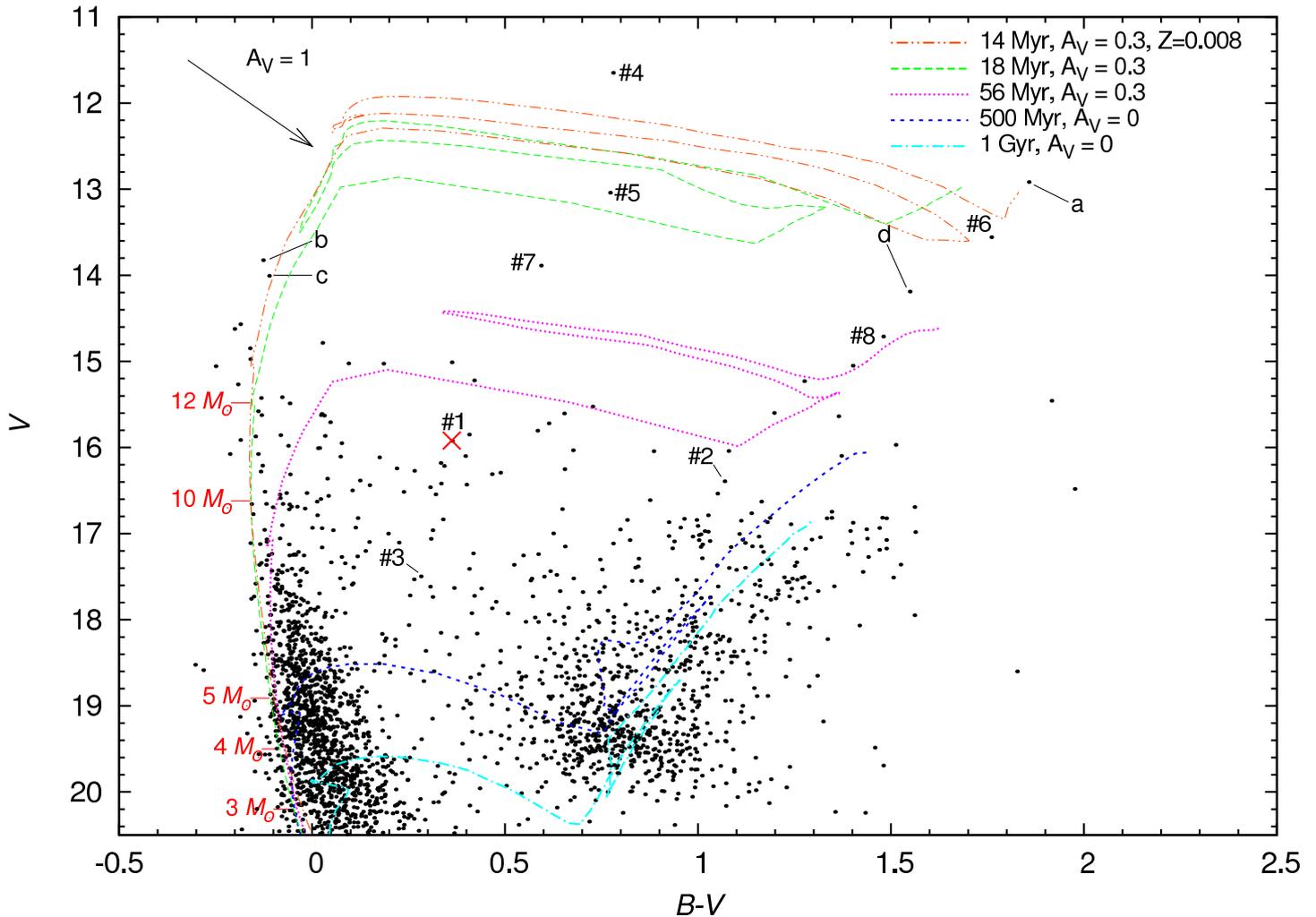}
\caption{Color-magnitude, {\it V} versus {\it B -- V}, diagram for 
stars observed toward SMC N33. 
Five  isochrones are shown, 18 Myr with $A_V = 0.3$ mag 
green dashed curve), 56 Myr (violet dotted curve), 
500 Myr (blue thick dashed curve),  1 Gyr 
(aqua dot-dashed curve), computed for a
metallicity of Z = 0.004 \citep{Lejeune01}, and 14 My 
(orange dash-double-dotted curve) for a metallicity of 
0.008 and a distance modulus of 18.94 mag.  
The cross indicates the location of the exciting star of N33. 
The numbers refer to the stars listed in Table\,\ref{tab:stars}. 
The letters $a$, $b$, $c$, and $d$ belong to the brightest stars of the photometry 
lying outside the field of Fig.\,\ref{fig:carte-champ}. They are located   
at positions  $\alpha$/$\delta$ = 
00:50:06.11/-73:28:12.06, 00:49:51.81/-73:25:16.49, 
00:50:03.74/-73:24:16.81, and 00:49:56.82/-73:28:26.17, respectively.     
}
\label{fig:hr}
\end{figure*}

\begin{figure*}[]
\centering
\includegraphics[width=13cm]{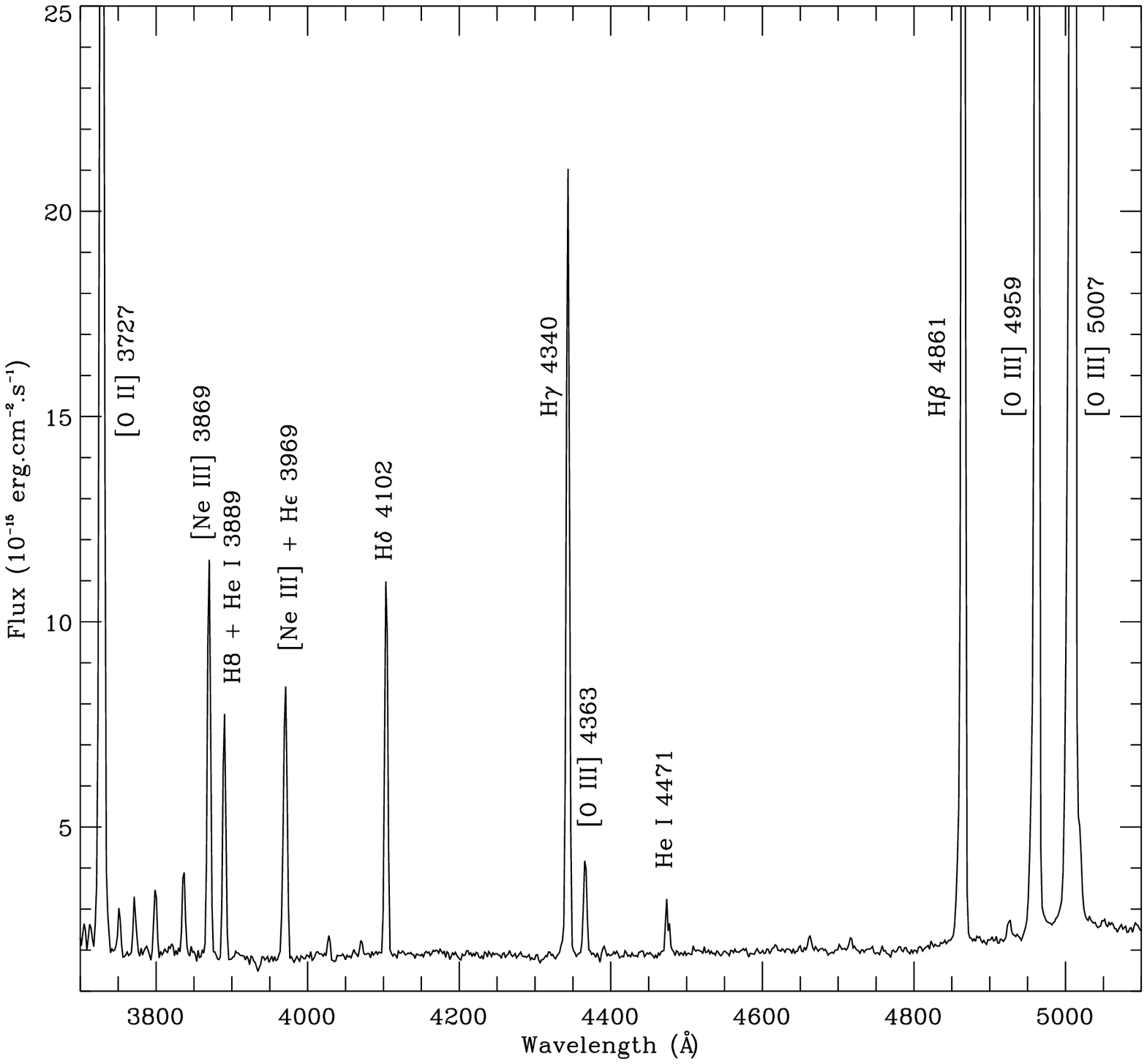}
\includegraphics[width=13cm]{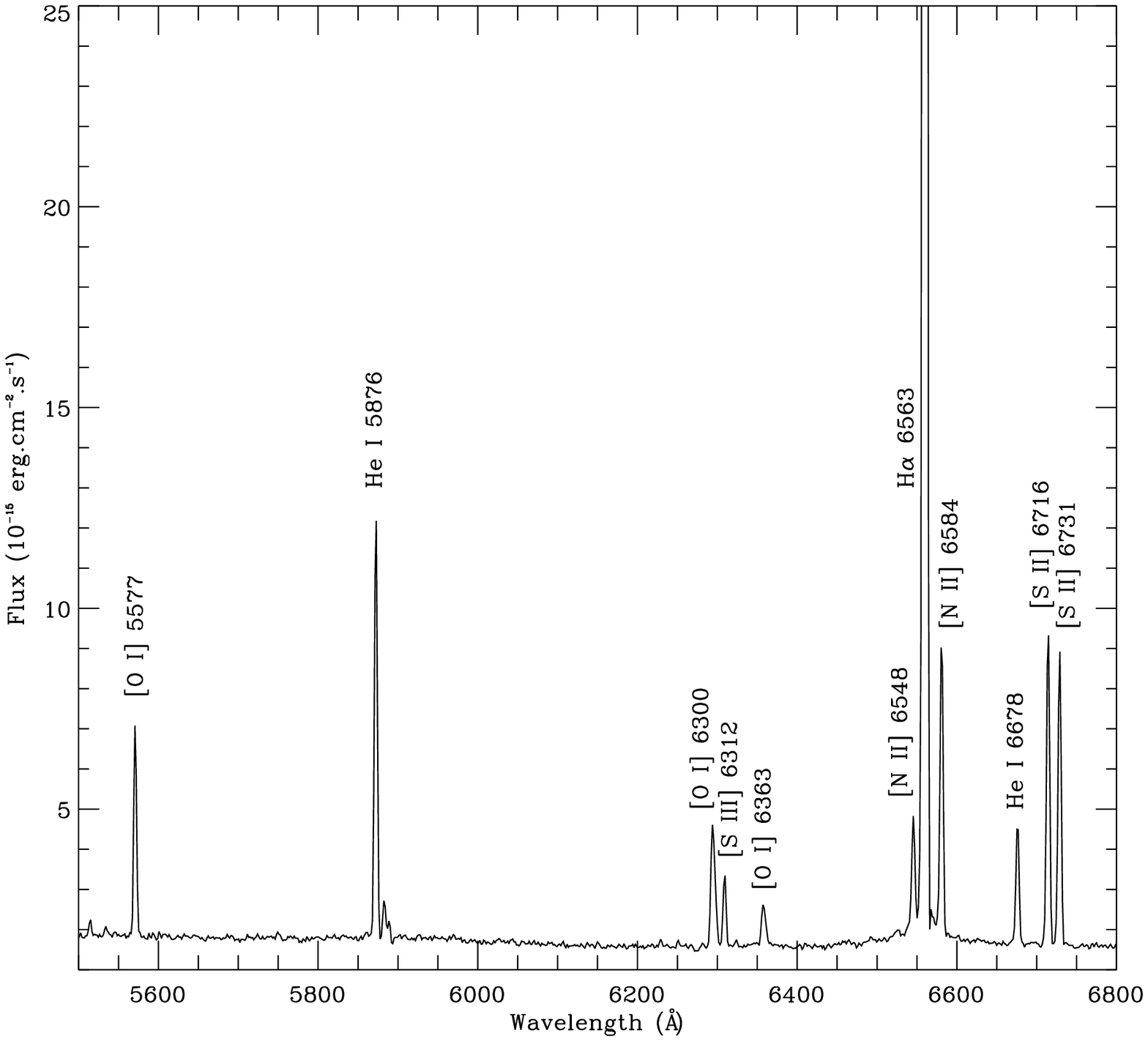}
\caption{Typical nebular spectra of the SMC compact \hii\ region N33 in the \hb\ (above) and 
\ha\ (below) regions.}
\label{fig:neb}
\end{figure*}

\begin{figure*}[]
\centering
\includegraphics[width=0.85\hsize]{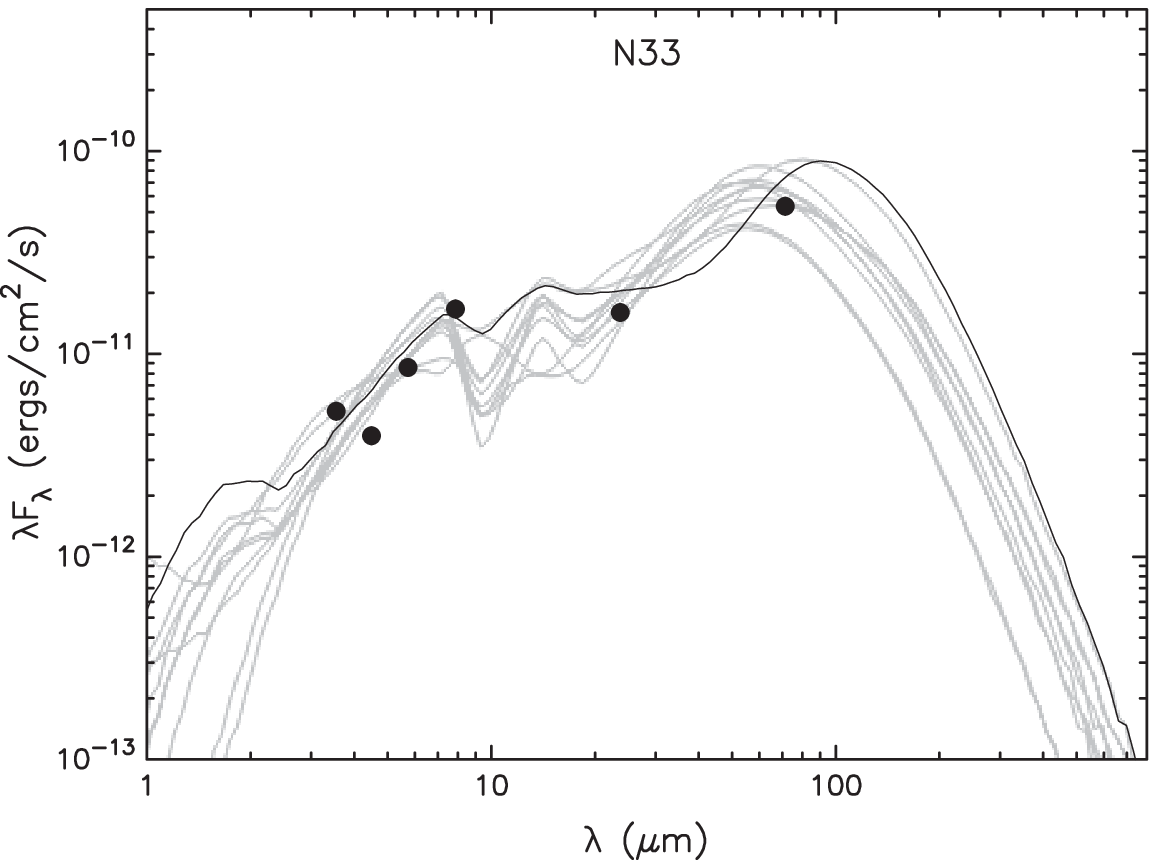}
\caption{
Mid-IR Spitzer photometry of the \hii\ region N33 fitted using 
YSO models \citep{Robitaille07}. 
The majority of the ten best-fit models suggest the presence of only 
envelope around the central stellar source of N33.  
The curve in black represents the best-fit. 
The derived total luminosity of the
nebula according to these models varies between 5\,\x\,10$^{3}$ and 10$^{4}$\,\slum. 
One should however be cautious in interpreting these results since N33 is not a YSO;  
see Sect. 3.8 for details.
}
\label{fig:sed}
\end{figure*}

\subsection{N33 as a HEB}

N33 should belong to the distinct and rare class of \hii\ 
regions in the Magellanic Clouds  called HEBs \citep[see][for a review]{MHM10b}.
In contrast to the typical \hii\ regions of 
the Magellanic Clouds, which are extended structures with sizes of several arc 
minutes corresponding to physical scales of more than 50\,pc and 
powered by a large number of exciting stars, HEBs are relatively dense and small 
regions of $\sim$\,5\frac\, to 10\frac\, in diameter in the optical, 
corresponding to $\sim$\,1.5 to 3.0\,pc  and excited by a much smaller number 
of massive stars. Their excitation, as derived from their  
\oiii\,(\lam \lam\,4959\,+\,5007)\,/\,\hb\ ratio, is generally 
larger than that of ordinary Magellanic Cloud \hii\ regions. 
For a fixed metallicity, the \oiii\,/\,\hb\ ratio increases with the effective 
temperature of the exciting star, 
as well as with the gas density in homogeneous photoionization models.\\

N33 qualifies for membership of this category. It is compact, with 
a mean angular radius of 3\frac.7, or 1.1 pc. The compactness implies a 
young age, since \hii\ regions disperse with time and become diffuse. 
It has a mean   \oiii\,/\,\hb\ ratio of 5.9, which is higher than 
the mean value for ordinary extended \hii\ regions in the Magellanic Clouds 
\citep[][and references therein]{MHM10b}. The  
electron temperature amounts to 12,540 K, which corresponds to 
excitation by a massive star at least as hot as an  
O6.5-O7 type at SMC metallicity. The high-excitation auroral 
transition of doubly ionized oxygen, \oiii\,\lam\,4363, is quite strong, as shown in 
Figs.\,\ref{fig:spectres} \& 8.    
Moreover, the estimated mean extinction over the \hii\ region, as derived 
in the optical, is  A$_{V}$ = 1.5 mag. This extinction peaks at 2.5 mag  
towards the northeast border, where the nebula is probably in contact with a 
molecular cloud of unknown characteristics. This association with a molecular 
cloud is another characteristic of HEBs. 
Moreover, in the plot comparing Magellanic Clouds compact \hii\ regions among them, 
N33 appears to be one of the most excited and brightest members of this class 
in the SMC \citep[][]{Meynadier07}.

\subsection{Spitzer SED fitting}

We used our Spitzer photometry to construct the mid-IR SED of N33 and
assess the behavior of the circumstellar dust of the nebula. We fitted
this SED with the library of YSO models by \cite{Robitaille06} using the
online SED fitting tool, provided by these authors 
\citep{Robitaille07}\footnote{ and available at 
http://caravan.astro.wisc.edu/protostars}.  
Figure\,\ref{fig:sed} shows the ten best-fit models to this SED. 
Only three of these models
include both circumstellar envelopes and disks. The majority of the
best-fit YSO models contain only one envelope around the
central stellar source of N33 with an outer radius varying between
3\,\x\,10$^{4}$~ and 10$^{5}$~AU. The derived total luminosity of the
nebula according to these models varies between 5\,\x\,10$^{3}$ to 10$^{4}$\,\slum. 
This agrees with the luminosity of \ab\,5\,\x\,10$^{4}$\,\slum,  
derived from IRAS 
fluxes considering that the IRAS SED peaks at \ab\,60\,$\mu$m.   
Unfortunately, the blue part of the complete SED of
N33, which includes the stellar source, could not be properly fitted.
There are four models available in the library of \cite{Robitaille06} that
include stellar sources with $T_{\rm eff}$,  corresponding to the
spectral type of the central star of N33  and a mass of
about 25\,\sm. However, their fit to the blue part of our SED was
not successful because there were few available measurements in the optical. 
In all cases, although SED fitting is an interesting exercise, its 
results should not be over-interpreted for the simple reason that N33 is not a YSO.

\section{Discussion}

N33 is quite noteworthy in that it is a isolated \hii\ region in this area of the 
SMC. In contrast to typical \hii\ regions in the Magellanic Clouds, 
which have neighboring \hii\ regions, N33 differs from any 
typical emission nebula. 
Thus it constitutes an interesting case for studies regarding the 
physical conditions in which massive stars form in isolation.
Observational findings suggest that massive star formation is a collective
process, i.e. that massive stars form in cluster environments 
\citep[e.g.,][]{Zinnecker07,Schilbach08,Gvaramadze08} 
and that the mass of the most massive star in a cluster is correlated with the 
mass of the cluster itself \citep[][]{Elmegreen00,Weidner06}. 
Nevertheless, other observational results imply that 
massive stars do not necessarily form in clusters but  
can be formed either as isolated stars or in very small groups. 
According to a statistical study by \citet[][]{deWit05}, nearly 95\% of Galactic O population 
is located in clusters or OB associations. This means that a small percentage, about 5\%, 
of high mass stars may form in isolation, in very good agreement with the 
finding by \citet[][]{Parker07}. 
Even in the Galactic center region, a number of Wolf-Rayet stars and O supergiants 
might have formed in isolation \citep[][]{Mauerhan10}. 
\citet[][]{Lamb10} observed eight apparently isolated O stars
in the SMC. Among the six non-runaway cases, there is no evidence of
clustering in three of them, which remain interesting candidates for
isolated massive-star formation. By isolation, we mean not clearly originating  
in an OB association. This definition therefore excludes runaway massive stars, 
which are thought to result from either dynamical interaction in massive dense clusters, 
or via a kick from a supernova explosion in a binary system   
\citep[e.g.,][and references therein]{Gvaramadze09,Pflamm-Altenburg10}.  
\citet[][]{Parker07} give a different definition for an isolated massive star:  
an O-type star belonging to a cluster whose total mass is $<$100\,\sm\, and 
moreover is devoid of B-type stars (10 $<M$/\sm  $<$ 17.5).
In the LMC, several massive runaway candidates have been reported 
\citep[e.g.][]{MHM00,Evans10,Gvaramadze10a}. Similarly, in the SMC a dozen OB 
runaway candidates were reported via detection of their bow shocks 
using Spitzer data \citep[][]{Gvaramadze10b}. \\

In any case, the massive star(s) powering N33 cannot be  runaway(s) for three main reasons: 
1) N33 is not linked to any OB association; 2) it seems impossible that 
a massive star carries its \hii\ region during the ejection, 
although ejection into a molecular cloud cannot be a priori excluded; and  
3) were N33 to contain more than one (probably exciting) stars, 
this situation would make the ejection hypothesis still less probable. 
On the other hand, the possibility that a massive star ejected into a molecular cloud 
creates an \hii\ region around 
itself has never been encountered. \\
 
To obtain a clear picture of how scattered the cold interstellar material 
is in the general area of N33, we investigate the distribution of molecular clouds 
and known young stellar objects (YSOs) in the vicinity ($\sim$~200~pc) of N33. 
CO surveys of the SMC reveal a concentration of 
molecular clouds in the southern end of this galaxy 
\citep[][]{Fukui10,Mizuno01,Stanimirovic00,Stanimirovic99,Blitz07,Bolatto07,Leroy07,
Bot10}. However, the available CO observations do not detect the  
N33 \hii\ region, maybe because of a lack of sensitivity or resolution  or for other 
reasons. The nearest molecular cloud to N33 is centered at a projected distance of 
\ab\,230 pc  \citep[][and references therein]{Fukui10}. As for YSOs,   
we used the catalog of candidates for these objects derived with
Spitzer from the S$^3$MC survey \citep{Bolatto07}. In this area, we found
five candidate YSOs, three of them being quite remote from N33 at
distances of between about 48 and 80~pc, hence we do not consider
them to be related to N33. On the other hand, we note
that the two remaining candidate YSOs, named S3MC\_J004930.12$-$732623.42
and S3MC\_J004929.07$-$732658.56 \citep[see][Table 4]{Bolatto07}, are
projected much closer to N33 at distances of about 5 and 7~pc, 
respectively (Fig.\,\ref{fig:spitzer}). The nature of these red 
sources as true YSOs still remains to be verified, 
and the possibility of a connection between N33 and these objects certainly 
requires further investigation, preferably with mid- and near-IR spectroscopy of these
sources. We also note that the S$^3$MC team did not identify N33 as a candidate YSO.  \\

It has been suggested that isolated field OB stars originate in a different 
mode of star formation from their counterparts in associations. 
\citet[][]{Massey02} and \citet[][]{Kroupa03} make the case that 
the field star initial mass function (IMF) might differ from that in 
OB associations/clusters. More explicitly, Galactic-field IMFs for early-type stars 
cannot, under any circumstances, be 
a Salpeter power-law, i.e. $\alpha$ = 2.35, but must have a steeper exponent,  
$\alpha$ $\geq$ 2.8  \citep[][]{Kroupa03}
Similarly, special physical conditions have been called up to account for  
the formation of very high-mass isolated stars in the bulge of the 
interacting Whirlpool galaxy, M51 \citep[][]{Lamers02}.
According to these authors, the formation of the bulge stars in M51 seems to be in line 
with theoretical predictions that isolated massive-star formation might take place in clouds 
where H$_{2}$, \oi\,63\,$\mu$m, and \cii\,158\,$\mu$m are the dominant coolants  
\citep[][]{Norman97,Mihos99}. These conditions are expected to occur in regions of 
low CO and dust contents because of the low metallicity, where  
the optical depth is \av\, $\leq$ 1 mag and the presence of  a hot source can dissociate 
the CO molecules \citep[][]{Lamers02}. 
The isolated massive stars might therefore be the result of these physical conditions  
and the luminous nucleus (situated at a distance between 90 and 270 pc).    
For comparison, the metallicity of SMC N33 is about a factor of 10 lower than solar. 
However, no major CO-dissociating hot source 
is present. The pre-eminent massive star cluster N66/NGC\,346 
\citep[][and references therein]{MHM10a} lies some 1400 pc away. Could this source 
have a similar effect as the M51 nucleus? Another similarity with M51 is 
that the SMC is also an interacting galaxy. \\

We note that in contrast to the above-mentioned proposal for the 
two different formation 
modes for cluster and isolated stars, \citet[][]{Oey04} 
find that the formation of field massive stars 
is a continuous process in associations. More specifically, 
the field stars do not originate from any different star-formation
mode. This conclusion is based on an empirical census of 
uniformly-selected massive-star candidates distributed all-over the SMC  
in more than 100 star clusters containing at least 3 to about 80 stars. 
Moreover, according to their Monte Carlo simulations, single, 
field OB stars are usually the most massive member of a group of smaller stars. 
This result is consistent with the idea that there is 
both a universal IMF and  universal $N_{*}^{-2}$ clustering law, which 
extends to $N_{*}$\,=\,1. Jointly, these laws imply that the fraction of field OB stars 
typically ranges from about 35\% to 7\% for most astrophysical 
situations.  \\

\citet[][]{Kauffmann10} suggest that massive stars might form only in 
molecular clouds possessing a minimum mass depending on the cloud size: 
{\it M(r)} $>$ 870 \sm\, ({\it r}/pc)$^{1.33}$. Thus, massive-star formation 
requires a large mass being concentrated in a relatively small volume. Clouds below this threshold 
might still form stars and clusters of up to intermediate mass (such as the Galactic Taurus, 
Ophiuchus, and Perseus molecular clouds). One can infer from this finding that 
molecular clouds with a mass slightly above this limit can form one or a couple of 
massive stars. N33 may represent such a case, but the problem is not at all  
solved, since it is then unclear why in this area of the SMC only one cloud 
happens to be massive enough to give rise to massive stars. In other words, we have yet to 
ascertain which physical conditions 
inhibit the aggregation of smaller clumps into massive clouds.    \\

In any case, the isolated massive star(s) powering N33 cannot be a result of   
the competitive accretion process \citep[][]{Bonnell97,Bonnell01}. 
According to this scenario, developing protostars in their natal molecular clouds 
compete with each other to gather mass. The protostars accrete mass with a rate
that depends on their location within the protocluster. They use the 
same reservoir of gas to grow. Hence, the  protostars nearest the
center, where the potential well is deep and gas densities are
higher, have the highest  accretion rates. This model is synonymous with 
collective massive-star formation. \\

There is no doubt that star \#1, the exciting source of the \hii\ region, is a 
massive star (of derived spectral type O6.5-O7\,V). We also note that this star 
constitutes the main object of the isolated massive-star formation debate presented in 
this paper. In contrast, the nature of the two neighboring stars \#2 ($V$ = 16.39, 
$B - V$= 1.07 mag) and \#3 ($V$ = 17.50, $B - V$ = 0.28 mag) is unclear. The presence of 
these stars raises several questions: Are they massive? Are they physically related to the 
compact \hii\, region and in particular to the massive star \#1? The present 
study cannot provide firm responses to these questions. 
We have found that {\bf if} there is a clustering in the region this occurs at 
a scale of about 3~pc and smaller, which actually covers the 1~pc separation between 
stars \#1, \#2, and \#3.  We note, however, that 
the color of star \#2 ($B - V$ = 1.07 mag) does not match that expected for a 
massive star assuming that stars \#1 and \#2 are more or less equally affected 
by reddening. As for star \#3, its absolute magnitude and color are compatible 
with a B-type star of mass about 15\,\sm. Further high-resolution observations, both 
imaging and spectroscopy, are necessary to probe the exact nature of this star 
and its relation to the \hii\ region.  \\
 
HEBs are usually located adjacent to ordinary giant \hii\ regions or seen lying 
across them. This implies that their formation is the consequence of triggering by 
a previous generation of massive stars in the complex 
\citep[][and references therein]{Elmegreen77,Whitworth94,Deharveng09}. 
However, this is not the case for N33 whose massive star(s) has(have) formed in isolation. 
Another example of an SMC HEB that has formed in isolation is N81, which was studied using 
{\it HST} imaging and spectroscopy \citep[][]{MHM99a,MHM02,Martins04}. 
It seems, however, that N33 is a comparatively more attractive 
candidate for isolated massive-star formation. N81 is an 
isolated \hii\ region in Shapley's wing. Nevertheless, it is certainly 
powered by at least two main O type stars, which are apparently part of 
a small cluster of massive stars detected with the {\it HST} spatial 
resolution \citep[][]{MHM99a}.     
The massive star at the origin of N33 has most certainly formed in an isolated, small 
molecular cloud, the mass and physical characteristics of which are not known. \\

It will be interesting to derive the IMF for such small 
massive star clusters formed in isolation, and compare it with that of massive stars 
formed in OB associations. We emphasize that individual studies of a variety of  
cases is essential in parallel with global, statistical investigations. 
These small clusters provide relatively simpler situations for 
studying the IMF because they involve a smaller number of physical factors 
because of their isolation. 
In the case of N33, it seems that the initial mass of the exciting star is \ab\,40\,\sm. 
It is unsurprising that the highest masses in small clusters be 
lower than those in OB associations. Indeed small clusters originate in molecular 
clouds with smaller masses. To advance the study of the mass distribution in small 
clusters, high resolution 
techniques in the optical and IR, both imaging and spectroscopy, environmental as well as  
stellar, are required. \\

\section{Concluding remarks}

This paper has presented the first detailed study of SMC N33 using imaging and spectroscopy 
in the optical obtained at the ESO NTT, as well as from the Spitzer and 2MASS 
data archives. We have derived a number of physical characteristics of N33 and 
ascertained its powering source. This compact \hii\ region of \ab\,7\frac.4 (2.2 pc) 
in diameter belongs to a small class of HEBs in the MCs. In contrast to other 
members of this class, N33 is not associated with any OB association. 
This object is excited by a  massive star of type O6.5-O7\,V, and represents an 
interesting case of isolated massive-star formation in the SMC.

\begin{acknowledgements}

We would like to thank the people in charge of Spitzer data archives, and 
particularly the PIs of the S$^{3}$MC and SAGE-SMC projects, 
Dr. Alberto D. Bolatto, University of Maryland, and Dr. Karl D. Gordon, 
University of Arizona, respectively. Our special thanks go also 
to the 2MASS facility for the data used in this study. 
D.A.G. kindly acknowledges financial support from the German Aerospace 
Center (DLR) through grant 50~OR~0908. We are indebted  
to Prof. Vassilis Charmandaris, Crete University, for his advice and help in 
using the Spitzer data. We would like also to thank Dr. Nolan R. Walborn, 
Space Telescope Institute, Baltimore, and Dr. Fabrice Martins, GRAAL, Montpellier, 
for commenting on a preliminary version of this paper. Last but not least, we are 
grateful to an anonymous referee for his meticulous review of the manuscript and several 
important suggestions that led to a much improved 
presentation of this research.  

\end{acknowledgements}


\begin{thebibliography}{}


\bibitem[Baldwin et al.(2000)]{Baldwin00} 
      Baldwin, J. A., Verner, E. M., Verner, D. A., et al. 2000, ApJS 129, 229


\bibitem[Battinelli(1991)]{battinelli91} 
	Battinelli, P.\ 1991, A\&A 244, 69 





\bibitem[Bica \& Schmitt(1995)]{Bica95} 
	Bica, E.~L.~D., \& Schmitt, H.~R.\ 1995, ApJS 101, 41 


\bibitem[Blitz et al.(2007)]{Blitz07}
     Blitz, L., Fukui, Y., Kawamura, A., et al. 2007, in Protostars and Planets V, 
        Reipurth, B. et al. (eds.), Univ. of Arizona Press, Tuscon, p. 81


\bibitem[Bolatto et al.(2007)]{Bolatto07} 
  Bolatto, A.~D., Simon, J. D., Stanimirovi\'c, S., et al.\ 2007, \apj\ 655, 212 


\bibitem[Bonnell et al.(1997)]{Bonnell97}
	Bonnell, I. A., Bate, M. R., Clarke, C. J., Pringle, J. E. 1997, 
	MNRAS 285, 201

\bibitem[Bonnell et al.(2001)]{Bonnell01} 
	Bonnell, I. A., Bate, M. R., Clarke, C. J., Pringle, J. E. 2001, 
	MNRAS 323, 785

\bibitem[Bot et al.(2010)]{Bot10}
	 Bot, C., Rubio, M., Boulanger, F. 2010, A\&A in press, astro-ph/1009.0124

\bibitem[Bouchet et al.(1985)]{Bouchet85}
	Bouchet, P., Lequeux, J., Maurice, E., et al. 1985 A\&A 149, 330



\bibitem[Castilho et al.(2000)]{Castilho00}
	Castilho, B. V., Gregorio-Hetem, J., Spite, F., et al. 2000, A\&A 364, 674 



\bibitem[Charmandaris et al.(2008)]{Charmandaris08} 
       Charmandaris, V., Heydari-Malayeri, M., Chatzopoulos, E. 2008, A\&A 396, 255 



\bibitem[Cignoni et al.(2009)]{Cignoni09}
	Cignoni1, M., Sabbi, E., Nota, A., et al. AJ  137, 3668   











\bibitem[Davies et al.(1976)]{DEM76}
      Davies, R. D., Elliott, K. H., Meaburn, J. 1976, MNRAS 81, 89 (DEM)


\bibitem[Deharveng et al.(2009)]{Deharveng09}
   	Deharveng, L., Zavagno, A., Schuller, F., et al. 2009, A\&A 496, 177

\bibitem[Dekker et al.(1986)]{Dekker86}
        Dekker, H., Delabre, B., Dodorico, S. 1986, SPIE 627, 339D

\bibitem[de Wit et al.(2005)]{deWit05}  
	de Wit, W. J., Testi, L., Palla, F., Zinnecker, H. 2005, A\&A 437, 247 

\bibitem[D'Odorico et al.(1998)]{D'Odorico98}
        D'Odorico, S., Beletic, J.W., Amico, P. et al. 1998, SPIE 3355, 507

\bibitem[Elmegreen \& Lada(1977)]{Elmegreen77} 
   Elmegreen, B.~G., Lada, C.~J. 1977, \apj\ 214, 725 

\bibitem[Elmegreen(2000)]{Elmegreen00} 
	Elmegreen  B. G. 2000, ApJ 539, 342 




\bibitem[Evans et al.(2010)]{Evans10} 	
      Evans, C.J., Walborn, N.R., Crowther, P.A., et al. 2010, ApJ 715, L74



\bibitem[Filipovi\'c et al.(1998)]{Filipovic98}
      Filipovi\'c, M. D., Haynes, R. F., White, G. L., Jones, P. A. 1998, A\&AS, 130, 421
   



\bibitem[Fukui \& Kawamura(2010)]{Fukui10}
	Fukui, Y., kawamura, A. 2010, ARAA 48, 547

\bibitem[Gomez et al.(1993)]{gomez93} 
	Gomez, M., Hartmann, L., Kenyon, S.~J., \& Hewett, R.\ 1993, AJ 105, 1927 

\bibitem[Gouliermis et al.(2000)]{Gouliermis00}
	Gouliermis, D., Kontizas, M., Korakitis, R., Morgan, D.~H., Kontizas, E., 
	Dapergolas, A. 2000, AJ 119, 1737
 





\bibitem[Gray \& Corbally(2009)]{Gray09}
    Gray, R.O., Corbally, C.J. 2009,  {\it Stellar Spectral Classification}, 
    Princeton Univ. Press

\bibitem[Gvaramadze \& Bomans(2008)]{Gvaramadze08} 
	Gvaramadze, V. V., Bomans, D. J. 2008, A\&A 490, 1071

\bibitem[Gvaramadze et al.(2009)]{Gvaramadze09} 
	Gvaramadze, V. V., Gualandris, A., Portegies Zwart, S. 2009, MNRAS 396, 570

\bibitem[Gvaramadze et al.(2010a)]{Gvaramadze10a}
	Gvaramadze, V. V., Kroupa, P., Pflamm-Altenburg, J. 2010a, A\&A, in press, 
        astro-ph/1006.0225

\bibitem[Gvaramadze et al.(2010b)]{Gvaramadze10b}
	Gvaramadze, V. V., Pflamm-Altenburg, J., Kroupa, P., 2010b, A\&A, in press, 
        astro-ph/1010.2490


\bibitem[Hatzidimitriou et al.(2005)]{Hatzidimitriou05} 
	Hatzidimitriou, D., Stanimirovi\'c, S., Maragoudaki, F., et al. 2005, MNRAS 360, 1171


\bibitem[Helou \& Walker(1988)]{Helou88} 
	Helou, G. \& Walker, D. W. 1988, {\it Infrared Astronomical Satellite 
        (IRAS) Catalogs and Atlases}, Vol. 7: The Small Scale Structure Catalog 
        (NASA RP-1190; Washington: GPO) 


\bibitem[Henize(1956)]{Henize56}
    Henize, K. G. 1956, ApJS 2, 315

\bibitem[Henize \& Westerlund(1963)]{Henize63}
     Henize, K. G., Westerlund, B. E. 1963, ApJ 137, 747











\bibitem[Heydari-Malayeri et al.(1999a)]{MHM99a} 
    Heydari-Malayeri, M., Rosa, M.~R., Zinnecker, H., Deharveng, L.,  Charmandaris, V.
   1999, A\&A 344, 848 

\bibitem[Heydari-Malayeri et al.(1999b)]{MHM99b} 
   Heydari-Malayeri, M.,  Charmandaris, V.,  Deharveng, L., Rosa, M.~R., Zinnecker, H.
    1999, \aap 347, 841 

\bibitem[Heydari-Malayeri et al.(2000)]{MHM00} 
	Heydari-Malayeri, M., Royer, P., Rauw, G., Walborn, N. R. 2000, A\&A 361, 877



\bibitem[Heydari-Malayeri et al.(2002)]{MHM02} 
   Heydari-Malayeri, M.,  Rosa, M.~R., Schaerer, D.,  Martins, F., Charmandaris, V.  
  2002, \aap 381, 951




\bibitem[Heydari-Malayeri \& Selier(2010)]{MHM10a} 
       Heydari-Malayeri, M., Selier, R. 2010, A\&A 517, A39    


\bibitem[Heydari-Malayeri et al.(2010)]{MHM10b} 
  Heydari-Malayeri, M., Rosa, M.~R.,  Charmandaris, V.,  et al. 2010, 
  {\it The Impact of HST on European Astronomy}, F. D. Macchetto (ed.), 
  Astrophysics and Space Science Proceedings, p. 31


\bibitem[Hodge(1985)]{Hodge85} 
	Hodge, P. 1985, PASP 97, 530

\bibitem[Howarth(1983)]{Howarth83}
    	Howarth, I. D. 1983, MNRAS 203, 301


\bibitem[Indebetouw et al.(2004)]{Indebetouw04}
	Indebetouw, R. Johnson, K. E., Conti, P. 2004, AJ 128, 2206


\bibitem[Jaschek \& Jaschek(1987)]{Jaschek87}
	Jaschek, C., Jaschek, M. 1987, The classification of stars, cambridge Univ. Press



\bibitem[Jacoby \& De Marco(2002)]{Jacoby02}
          Jacoby, G. H.,  De Marco, O. 2002, AJ, 123, 269

\bibitem[Johnson et al.(2006)]{Johnson06} 
    Johnson, M. D., Levitt, J. S., Henry, R. B. C., et al. 2006, Proceedings IAU Sympo. 234, 
    {\it Planetary Nebulae in Our Galaxy and Beyond}, 
    M.J. Barlow \& R.H. M\'endez (eds.), astro-ph/0605099


\bibitem[Kauffmann \& Pillai(2010)]{Kauffmann10}
	Kauffmann, J., Pillai, T. 2010, astro-ph/1009.1617, ApJ Letters, in press 





\bibitem[Koornneef(1983)]{Koornneef83} 
	Koornneef, J. 1983, A\&A 128, 84


\bibitem[Kroupa \& Weidner(2003)]{Kroupa03} 	
	Kroupa, P., Weidner, C. 2003, ApJ 598, 1076

\bibitem[Kwitter \& Henry(2001)]{Kwitter01} 
      	Kwitter, K. B., Henry, R. B. C. 2001, ApJ 562, 804

\bibitem[Lindsay(1961)]{Lindsay61}
         Lindsay, E. M. 1961, AJ 66, 169   

\bibitem[Lamb et al.(2010)]{Lamb10}
        Lamb, J. B., Oey,  M. S., Werk, J. K., Ingleby, L. D. 2010, 
        ApJ, in press, astro-ph/1010.5273


\bibitem[Lamers et al.(2002)]{Lamers02} 
	Lamers, H. J. G. L. M., Panagia, N., Scuderi, S., et al. 2002, ApJ 566, 818

\bibitem[Lamers(2005)]{Lamers05}
        Lamers, H. J. G. L. M. 2005, Proc. IAU Symp. 227, Cesaroni et al. (eds.),  
        Cambridge Univ. Press, 303

\bibitem[Landolt(1992)]{Landolt92}
	Landolt, A. U. 1992, AJ 104, 340 


\bibitem[Laney \& Stobie(1994)]{Laney94} 
	Laney, C. D., Stobie, R. S. 1994, MNRAS 266, 441


\bibitem[Larson(1995)]{larson95} 
	Larson, R.~B.\ 1995, MNRAS 272, 213 




\bibitem[Lee et al.(2009)]{Lee09}
	Lee, M.-Y., Stanimirovi\'c, S., J\"{u}rgen, O., et al. 2009, AJ 138, 1101    


\bibitem[Lejeune \& Schaerer(2001)]{Lejeune01}
	Lejeune, T., Schaerer, D. 2001, A\&A 366, 538 

\bibitem[Leroy et al.(2007)]{Leroy07}
     Leroy, A., Bolatto, A., Stanimirovi\'c, S., et al. 2007, ApJ 658, 1027

\bibitem[Levesque et al.(2006)]{Levesque06}
	Levesque, E.M., Massey, P., Olsen, K. A. G. et al. 2006, ApJ 645, 1102


\bibitem[Martins et al.(2004)]{Martins04}
	Martins, F., Schaerer, D., Hillier, D. J., Heydari-Malayeri, M. 2004, A\&A 420, 1087

\bibitem[Martins et al.(2005)]{Martins05}
	Martins, F., Schaerer, D., Hillier, D. J. 2005, \aap\ 436, 1049    

\bibitem[Martins \& Plez(2006)]{Martins06}
	Martins, F., Plez, B. 2006, A\&A 457, 637 

\bibitem[Massey (2002)]{Massey02}
	Massey, P. 2002, ApJS 141, 81



\bibitem[Mathewson et al.(1986)]{Mathewson86}
        Mathewson, D. S., Ford, V. L., Visvanathan, N. 1986, ApJ 301, 664 

\bibitem[Mathewson et al.(1988)]{Mathewson88}
        Mathewson, D. S., Ford, V. L., Visvanathan, N. 1988, ApJ 333, 617

\bibitem[Mauerhan et al.(2010)]{Mauerhan10}
        Mauerhan, J., Cotera, A., Dong, H., et al. 2010, accepted to ApJS, astro-ph/1009.2769

\bibitem[McGee \& Newton(1981)]{McGee81}
	McGee, R. X., Newton, L. M. 1981, PASAu 4, 189


\bibitem[McGee \& Newton(1982)]{McGee82} 
	McGee, R. X., Newton, L. M. 1982, PASAu 4, 308





\bibitem[Meyssonnier \& Azzopardi(1993)]{Azzo93}
    	Meyssonnier, N., Azzopardi, M. 1993, A\&AS 102, 451

\bibitem[Meynadier \& Heydari-Malayeri(2007)]{Meynadier07} 
	Meynadier, F., \& Heydari-Malayeri, M.\ 2007, \aap, 461, 565


\bibitem[Mihos et al.(1999)]{Mihos99}
	Mihos, J. C., Spaans, M., McGaugh, S. S. 1999, ApJ 515, 89


\bibitem[Mizuno et al.(2001)]{Mizuno01}
	Mizuno, N., Rubio, M., Mizuno, A., et al. 2001, PASJ 53, L45


   




\bibitem[Norman \& Spaans(1997)]{Norman97}
	Norman, C. A., Spaans, M. 1997, ApJ 480, 145



\bibitem[Oey et al.(2004)]{Oey04}
	Oey, M. S., King, N. L., Parker, J. Wm. 2004, AJ 127, 1632

\bibitem[Parker \& Goodwin(2007)]{Parker07}
        Parker, R. J., Goodwin, S. P. 2007, MNRAS 380, 1271


\bibitem[Peimbert \& Costero(1969)]{Peimbert69}
	Peimbert, M., Costero, R. 1969, Bol. Obs. Tonantzintla y Tacubaya, 5, 3



\bibitem[Pflamm-Altenburg \& Kroupa(2010)]{Pflamm-Altenburg10}
	Pflamm-Altenburg, J., Kroupa, P. 2010, MNRAS 404, 1564

\bibitem[Pr\'evot et al.(1984)]{Prevot84}
	Pr\'evot, M. L., Lequeux, J., Prevot, L., et al. 1984, A\&A 132, 389 


  

\bibitem[Robitaille et al.(2006)]{Robitaille06}
	Robitaille, T.~P., Whitney, B.~A., Indebetouw, R., et al. 2006, \apjs 167, 256


\bibitem[Robitaille et al.(2007)]{Robitaille07} 
	Robitaille, T.~P.,Whitney, B.~A., Indebetouw, R., \& Wood, K. 2007, 
	\apjs 169, 328



\bibitem[Russell \& Dopita(1992)]{Russell92}
	Russell, S. C., Dopita, M. A. 1992, ApJ 384, 508




\bibitem[Schilbach \& R\"oser(2008)]{Schilbach08}
	Schilbach, E., R\"oser, S.  2008, A\&A 489, 105

\bibitem[Shapley(1940)]{Shapley40}    
    Shapley H. 1940, BHarO 914, 8

\bibitem[Shaw \& Dufour(1995)]{Shaw95}
	Shaw, R. A., Dufour, R. J. 1995, PASP, 107, 896






\bibitem[Stanimirovi\'c et al.(1999)]{Stanimirovic99}
    Stanimirovi\'c, S., Staveley-Smith, L., Dickey, J. M. et al. 1999, MNRAS 302, 417

\bibitem[Stanimirovi\'c et al.(2000)]{Stanimirovic00}
    	Stanimirovi\'c, S., Staveley-Smith, L., van der Hulst, J. M., et al. 2000, MNRAS 315, 791

\bibitem[Staveley-Smith et al.(1997)]{Staveley-Smith97} 
   Staveley-Smith, L., Sault, R. J., Hatzidimitriou, et al. 1997, MNRAS 289, 225 
















\bibitem[Weidner \& Kroupa(2006)]{Weidner06} 
	Weidner C., Kroupa  P. 2006, MNRAS 365, 1333
 

\bibitem[Westerlund(1990)]{Westerlund90}
	Westerlund, B. E. 1990, A\&ARv 2, 29


\bibitem[Whitworth et al.(1994)]{Whitworth94} 
    Whitworth, A. P., Bhattal, A. S., Chapman, S. J., et al. 1994, MNRAS 268, 291

\bibitem[Wilke et al.(2003)]{Wilke03} 
    Wilke, K., Stickel, M., Haas, M., et al. 2003, A\&A 401, 873




 



\bibitem[Zinnecker \& Yorke(2007)]{Zinnecker07} 
  Zinnecker, H., \& Yorke, H.~W.\ 2007, \araa, 45, 481 
 
\end{thebibliography}
\end{document}